\documentclass[prd,twocolumn,showpacs,preprintnumbers,nofootinbib,amsmath,eqsecnum]{revtex4}

 \usepackage[dvips,final]{graphicx}
  \usepackage{amssymb}
   \usepackage{amsmath}
    \usepackage{amsfonts}
     \usepackage{epsfig}
      \usepackage{bm} % bold math
\usepackage{mathrsfs}
\usepackage{slashed}      
      
\usepackage[colorlinks = true,
linkcolor = blue,
urlcolor  = blue,
citecolor = blue,
anchorcolor = blue]{hyperref}

\begin{document}

\title{Gravitational chiral anomaly for spin 3/2 field interacting with spin 1/2 field}

\author{G. Yu. Prokhorov}
\email{prokhorov@theor.jinr.ru}
\affiliation{Joint Institute for Nuclear Research, Joliot-Curie str. 6, Dubna 141980, Russia}
\affiliation{Institute of Theoretical and Experimental Physics, NRC Kurchatov Institute, 
B. Cheremushkinskaya 25, Moscow 117218, Russia}
\author{O. V. Teryaev}
\email{teryaev@jinr.ru}
\affiliation{Joint Institute for Nuclear Research, Joliot-Curie str. 6, Dubna 141980, Russia}
\affiliation{Institute of Theoretical and Experimental Physics, NRC Kurchatov Institute, 
	B. Cheremushkinskaya 25, Moscow 117218, Russia}
\author{V. I. Zakharov}
\email{vzakharov@itep.ru}
\affiliation{Institute of Theoretical and Experimental Physics, NRC Kurchatov Institute, 
B. Cheremushkinskaya 25, Moscow 117218, Russia}
\affiliation{Pacific Quantum Center, 
Far Eastern Federal University, 10 Ajax Bay, Russky Island, Vladivostok 690950, Russia\vspace{0.8 cm}}

\begin{abstract}
The gravitational chiral quantum anomaly is calculated in the framework of the extended Rarita-Schwinger-Adler (RSA) field theory, which includes the interaction with an additional spin 1/2 field. It is shown that the factor in the gravitational chiral anomaly, normalized to the Dirac field anomaly, is equal to --19. The resulting value distinguishes the RSA theory from the other theories of spin 3/2.
A direct verification of the conformality of the RSA theory in the strong interaction limit at the level of one-loop three-point graphs was made.
\end{abstract}

\maketitle

%================
\section{Introduction}
\label{sec_intro}
%================

Quantum anomalies for fields with spin 1/2 have been well studied over the past years. In particular, there is a famous axial quantum anomaly, which includes gauge and gravitational (also called mixed axial-gravitational) parts
\begin{eqnarray}
\langle \nabla_{\mu}\hat{j}^{\mu}_A\rangle_{S=1/2}&=&-\frac{1}{16\pi^2 \sqrt{-g}}\varepsilon^{\mu\nu\alpha\beta}F_{\mu\nu}F_{\alpha\beta} \nonumber \\
&&+\frac{1}{384 \pi^2\sqrt{-g}}\varepsilon^{\mu\nu\rho\sigma}
R_{\mu\nu\kappa\lambda}{R_{\rho\sigma}}^{\kappa\lambda}\,,\,\,\,
\label{anom12}
\end{eqnarray}
where $ F_{\mu\nu} $ is the gauge field strength tensor, $ R_{\mu\nu\kappa\lambda} $ is the curvature tensor and $ \nabla_{\mu} $ is the covariant derivative.

In the case of higher spins, the theory turns out to be more complicated and, as a rule, has internal problems, which makes the question of quantum anomalies not completely transparent.

Recently, a new value was obtained for the gauge chiral anomaly in the case of spin 3/2 \cite{Adler:2017shl, Adler:2019zxx}. One of the most common ways of constructing a spin 3/2 theory is based on the Rarita-Schwinger field. However, the conventional Rarita-Schwinger field theory has a number of pathologies (see \cite{Adler:2015yha} and references therein). In \cite{Adler:2017shl} the extended Rarita-Schwinger-Adler (RSA) theory was proposed, in which some of the pathologies were overcome by introducing a nontrivial chirally symmetric interaction with an additional spin 1/2 field.

% In particular, the problem with a jump in the number of degrees of freedom was solved when the interaction with the gauge field was turned on.

In particular, an important problem with the singularity in the Dirac bracket in the limit of weak gauge fields was solved, which allowed to consistently gauge the theory beyond the supergravity approach. Since the Rarita-Schwinger field theory is a generalized Hamiltonian dynamics, the quantum anticommutator is given by the Dirac bracket, not the Poisson bracket. The existence of a singularity in this bracket does not allow constructing a perturbation theory. The introduction of the interaction with an additional spin 1/2 field in \cite{Adler:2017shl} shifted the pole in the Dirac bracket, which, in turn, allowed to find the gauge chiral anomaly by the famous shift method
\begin{eqnarray}
\langle\partial_{\mu}\hat{j}^{\mu}_A\rangle_{RSA}=-\frac{5}{16\pi^2}\varepsilon^{\mu\nu\alpha\beta}F_{\mu\nu}F_{\alpha\beta}\,.
\label{anom}
\end{eqnarray}
Thus, the gauge chiral anomaly turns out to be 5 times larger than for spin 1/2. This numerical factor distinguishes RSA theory from other approaches to the description of Rarita-Schwinger fields, such as supergravity \cite{Duff:1982yw}.

Subsequently, the anomaly (\ref{anom}) was generalized to the non-Abelian case and used for the grand unification in \cite{Adler:2019exo}, where fields with spin 3/2 participate in the anomaly cancellation.

In this paper, we continue to analyze the RSA theory and find the gravitational chiral anomaly. For Rarita-Schwinger fields, there is a well-known result for the gravitational chiral anomaly \cite{Duff:1982yw}
\begin{eqnarray}
\langle\nabla_{\mu} \hat{j}^{\mu}_A\rangle_{RS}=
\frac{-21}{384 \pi^2\sqrt{-g}}\varepsilon^{\mu\nu\rho\sigma}
R_{\mu\nu\kappa\lambda}{R_{\rho\sigma}}^{\kappa\lambda}\,,
\label{anom_old}
\end{eqnarray}
where the factor in the anomaly is $-21$ times larger compared to the anomaly for Dirac field. This value has been obtained in various ways, in particular, in supergravity \cite{Duff:1982yw, Carrasco:2013ypa}, as well as for quantized Rarita-Schwinger fields on a classical geometric background \cite{Nielsen:1978ex} (with restrictions imposed on the geometry). Note that (\ref{anom12}) and (\ref{anom_old}) are special cases of the more general relation, obtained in the framework of the supergravity \cite{Duff:1982yw}, for an arbitrary spin $ S $
\begin{eqnarray}
\langle\nabla_{\mu} \hat{j}^{\mu}_A\rangle_S =
\frac{(S-2S^3)}{96 \pi^2\sqrt{-g}}\epsilon^{\mu\nu\rho\sigma}
R_{\mu\nu\kappa\lambda}{R_{\rho\sigma}}^{\kappa\lambda}\,.
\label{anom_gen}
\end{eqnarray}

Already in the case of the gauge anomaly (\ref{anom}), the result differed from the other theories with Rarita-Schwinger field. Therefore, we expect in advance that the RSA theory will give a new numerical factor also for the gravitational chiral anomaly instead of (\ref{anom_old}). The study of this issue is the aim of this work.

\begin{figure}[h!]
	\begin{center}
		\includegraphics[width=0.45\textwidth]{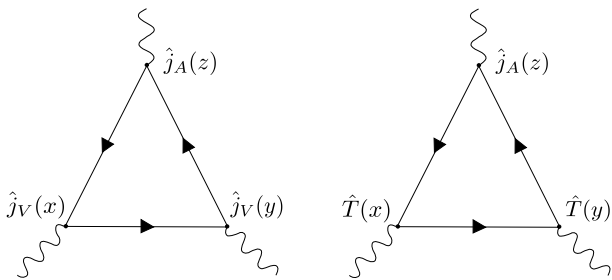}
	\end{center}
	\caption{\label{fig}Typical diagrams to be calculated in  \cite{Erdmenger:1996yc,Erdmenger:1999xx} to find the gauge chiral anomaly (left) and the gravitational chiral anomaly (right).}
\end{figure}

To find the anomalies, we use a new method described in \cite{Erdmenger:1996yc, Erdmenger:1999xx} and based on considering the form of the three-point quantum correlation function. In this case a three-point function with two vector current operators and one axial current operator defines a gauge chiral anomaly. A three-point function with two stress-energy tensors and an axial current defines a gravitational chiral anomaly. Typical diagrams to be found are shown in Fig.~\ref{fig}. 

It was shown that in conformally symmetric theories these three-point functions should have a universal form. The specific choice of the theory affects only the numerical factor in front of the universal function and the key point is the equality of this numerical factor to the coefficient in the quantum anomalies: gauge and gravitational.

The advantage of this method for calculating anomalies is that all calculations can be made in a coordinate representation and, thus, we do not need explicit taking non-trivial divergent loop  momentum integrals. Also, in the case of a gravitational chiral anomaly, the correlator (Fig. \ref{fig} on the right), can be found in a flat space-time and we do not need to consider Rarita-Schwinger fields in curved space. This method of finding anomalies corresponds to 't Hooft's interpretation, according to which anomalies are properties of the quantum theory, and gauge or gravitational fields only make them visible in conservation laws \cite{Chernodub:2021nff}.

First, we reproduce the gauge quantum anomaly (\ref{anom}): this allows us to demonstrate the features of the method. Then we move on to a more complicated case and calculate the gravitational chiral anomaly in RSA theory.

Our result for the anomaly can be further used in the extended, beyond the Standard model, field theories containing higher spin fields for the cancellation of the gravitational chiral anomalies. In particular, it can be considered in the framework of the theory \cite{Adler:2019exo}.

But additional motivation for us comes from a new field, considering the manifestations of quantum anomalies in relativistic fluids and condensed matter. Namely, there is a new phenomenon, the chiral vortical effect (CVE) \cite{Son:2009tf,Zakharov:2012vv, Kharzeev:2013jha}, which is a transfer of chirality along the vorticity in a vortical fluid.

This effect turns out to be related to the gauge chiral anomaly \cite{Son:2009tf, Sadofyev:2010is, Zakharov:2012vv}, which was explicitly shown for the case of spin 1/2. The case of higher spins is less obvious, but in \cite{Prokhorov:2021bbv} we derived the CVE in the RSA theory and showed its relationship with the anomaly (\ref{anom}). We also note, that similar relationship was substantiated for the RSA theory for a close phenomenon, chiral separation effect (CSE), in the presence of a magnetic field \cite{Khaidukov:2020mrn}.

Much less trivial is the question of manifestations of the gravitational chiral anomaly \cite{Stone:2018zel, Landsteiner:2011cp, Prokhorov:2020okl, Prokhorov:2020npf, Chernodub:2021nff}, and is based on the interpretation of Hawking radiation as an effect of the anomaly on the black hole horizon \cite {Robinson:2005pd}. The calculation of the axial-gravitational anomaly for the RSA theory will allow us to analyze the existing concepts of the relationship between the gravitational chiral anomaly and thermodynamics in the case of higher spins.

The paper has the following structure. In Section \ref{sec_theory} we give short introduction into the extended Rarita-Schwinger-Adler theory. In Section \ref{sec_method} we describe a method for calculating quantum anomalies based on an analysis of the form of three-point functions. Section \ref{sec_calc} contains original results: the derivation of the gauge chiral anomaly and the gravitational chiral anomaly in the RSA theory. Section \ref{sec_disc} provides an interpretation of the obtained factor in the gravitational chiral anomaly. The Conclusion lists the main results. Appendix \ref{append} contains the calculation of the three-point function needed to find the gauge chiral anomaly, in a general case of arbitrary points.

Analytical calculations in Section \ref{sec_calc} and Appendix \ref{append} were made using the Wolfram Mathematica system for technical computing and the package \cite{Patel:2015tea}, on the parallel computing server Theor4 of JINR BLTP.

We use the notations $\eta_{\mu\nu} = \mathrm{diag} (1, -1, -1, -1)$, $\epsilon^{0123}=1$, $ \gamma_5=i\gamma^0\gamma^1\gamma^2\gamma^3 $, and the system of units $ e=\hbar=c=1 $.

%================
\section{The theory of spin 3/2 field interacting with a field with spin 1/2}
\label{sec_theory}
%================

In \cite{Adler:2017shl} S. L. Adler formulated an extended theory for spin 3/2.
It is described by an action of the form
\begin{eqnarray}
S&=&\int d^4 x\,\big( -\varepsilon^{\lambda \rho \mu \nu} \bar{\psi}_{\lambda}\gamma_5 \gamma_{\mu} \partial_{\nu} \psi_{\rho}+i \bar{\lambda} \gamma^{\mu}\partial_{\mu} \lambda \nonumber \\
&&- i m \bar{\lambda} \gamma^{\mu}\psi_{\mu}+i m \bar{\psi}_{\mu} \gamma^{\mu}\lambda \big)\,,
\label{action}
\end{eqnarray}
where $ \psi_{\mu} $ is the Rarita-Schwinger field, $ \lambda $ is an additional field with spin 1/2, and $ m $ is the interaction constant. These interaction terms have the form of ``proto-massive'' terms: they describe the interaction of the fields $ \psi_{\mu} $ and $ \lambda $ with the same-sign chirality, which does not allow us to consider them as usual massive terms. Following \cite{Adler:2017shl}, we will consider the limit $ m\to \infty $. In this limit negative Dirac bracket vanishes, and also the calculations become more compact.

The introduced interaction with the $ \lambda$ field shifts the pole in the Dirac bracket in the presence of an external gauge field, which ultimately made it possible to find the chiral quantum anomaly.

In some cases, it is convenient to write the action (\ref{action}) in terms of the product of three Dirac matrices
\begin{eqnarray}
\gamma^{\lambda\rho\nu} &=& \frac{1}{6}(\gamma^{\lambda}\gamma^{\rho}\gamma^{\nu}-\gamma^{\lambda}\gamma^{\nu}\gamma^{\rho}-\gamma^{\rho}\gamma^{\lambda}\gamma^{\nu}+\gamma^{\rho}\gamma^{\nu}\gamma^{\lambda} \nonumber \\
&&+\gamma^{\nu}\gamma^{\lambda}\gamma^{\rho}-\gamma^{\nu}\gamma^{\rho}\gamma^{\lambda})= \nonumber \\
&=& \frac{1}{2}(\gamma^{\lambda}\gamma^{\rho}\gamma^{\nu}-\gamma^{\nu}\gamma^{\rho}\gamma^{\lambda})= \nonumber \\
&=& i\varepsilon^{\lambda \rho \mu \nu}\gamma_5\gamma_{\mu} \,.
\label{ggg}
\end{eqnarray}

The classical equations of motion can be obtained by varying the fields $ \psi_{\mu} $ and $ \lambda $ and have the form
\begin{eqnarray}
\varepsilon^{\lambda \mu \nu \rho}\gamma_5 \gamma_{\mu} \partial_{\nu} \psi_{\rho} - i m \gamma^{\lambda} \lambda&=&0\,,  \nonumber\\
\varepsilon^{\rho \nu \lambda \mu}\partial_{\nu} \bar{\psi}_{\lambda} \gamma_5 \gamma_{\mu} + i m \bar{\lambda} \gamma^{\rho} &=&0\,,  \nonumber\\
\gamma^{\mu}\partial_{\mu} \lambda- m \gamma^{\mu} \psi_{\mu} &=&0\,,  \nonumber\\
\partial_{\mu} \bar{\lambda} \gamma^{\mu} - m \bar{\psi}_{\mu}\gamma^{\mu}  &=&0\,.
\label{eq1}
\end{eqnarray}
Taking the derivative in the first pair of equations and using the second pair of equations, we obtain
\begin{eqnarray}
\gamma_{\mu} \psi^{\mu}  =\bar{\psi}^{\mu} \gamma_{\mu}   =
\gamma_{\mu} \partial^{\mu} \lambda  = 
\partial^{\mu} \bar{\lambda}  \gamma_{\mu}=0\,.
\label{eq2}
\end{eqnarray}

To derive the stress-energy tensor, it is necessary to pass to an arbitrary space-time metric
\begin{eqnarray}
S&=& \int d^4 x \sqrt{-g}\bigg(\frac{i}{2}\Big[\bar{\psi}_{\lambda} \gamma^{\lambda\rho\nu} D_{\nu} \psi_{\rho}-
D_{\nu}\bar{\psi}_{\lambda}\gamma^{\lambda\rho\nu}\psi_{\rho}\Big]   \nonumber \\ &&+\frac{i}{2}\Big[\bar{\lambda}\gamma^{\mu}D_{\mu}\lambda-
D_{\mu}\bar{\lambda}\gamma^{\mu}\lambda\Big]\nonumber \\ &&+
i\, m \Big[\bar{\psi}_{\mu}\gamma^{\mu}\lambda 
- \bar{\lambda}\gamma^{\mu}\psi_{\mu}\Big]\bigg)\,.
\label{action_curved}
\end{eqnarray}
using tetrads $ e^{\mu}_a $. Tetrads satisfy the usual relations
\begin{eqnarray}
&&\gamma^{\mu}=e^{\mu}_a \gamma^a\,,\quad
e^a_{\mu}e_{a\nu}=g_{\mu\nu}\,, \nonumber \\
&&e^{\mu}_{a}e_{b\mu}=\eta_{ab}=(1,-1,-1,-1)
\,.
\label{tetrads}
\end{eqnarray}
The covariant derivative for the Rarita-Schwinger field contains a term with Christoffel symbols $ \Gamma^{\sigma}_{\nu\rho} $ and a spinor connection $ \omega^{ab}_{\nu} $
\begin{eqnarray} \label{cov_der}
D_{\nu} \psi_{\rho} &=&\partial_{\nu} \psi_{\rho} -
\Gamma^{\sigma}_{\nu\rho}\psi_{\sigma}+\frac{i}{2}\omega^{ab}_{\nu} \sigma_{ab}\psi_{\rho}\,, \\
D_{\nu} \lambda &=&\partial_{\nu} \lambda +\frac{i}{2}\omega^{ab}_{\nu} \sigma_{ab}\psi_{\rho}
\,,\nonumber \\
\sigma_{ab}&=&\frac{i}{2}[\gamma^a,\gamma^b]
\,,\nonumber \\
\Gamma^{\sigma}_{\nu\rho}&=&\frac{1}{2}g^{\sigma\alpha}(\partial_{\nu}g_{\alpha\rho}
+\partial_{\rho}g_{\alpha\nu}
-\partial_{\alpha}g_{\nu\rho})
\,,\nonumber \\
\omega^{ab}_{\nu}&=&\frac{1}{4}(e^{b\lambda}\partial_{\nu}e^a_{\lambda}-e^{a\lambda}\partial_{\nu}e^{b}_{\lambda})+\frac{1}{4}(e^b_{\tau}e^{a\lambda}-e^a_{\tau}e^{b\lambda})\Gamma^{\tau}_{\lambda\nu}
\,.\nonumber 
\end{eqnarray}
The stress-energy tensor can be obtained by varying with respect to the metric
\begin{eqnarray}
T^{\mu\nu}=-\frac{2}{\sqrt{-g}}\frac{\delta S}{\delta g_{\mu\nu}} \,.
\label{temvar}
\end{eqnarray}
It is convenient to first vary all the quantities with respect to the metric $ g'_{\mu\nu}=\eta_{\mu\nu}+h_{\mu\nu} $ (see, e.g. \cite{Shapiro:2016pfm})
\begin{eqnarray}
&&\delta g^{\mu\nu}=-h^{\mu\nu}\,,\quad
\delta e_a^{\mu}=-\frac{1}{2} h^{\mu}_{\rho}e_a^{\rho}\,,
\nonumber \\
&&
\delta \sqrt{-g}=\frac{1}{2}\sqrt{-g}h^{\mu}_{\mu}\,,\quad \delta \gamma^{\mu}=-\frac{1}{2} h^{\mu}_{\rho}\gamma^{\rho}\,,\nonumber \\
&&
\delta \omega^{ab}_{\mu}=-\frac{1}{4} (e^{a\tau}e^{b\lambda}-e^{b\tau}e^{a\lambda})D_{\lambda} h_{\mu\tau}\,,
\label{vary_curved}
\end{eqnarray}
where the above relations are valid in the first order in $ h_{\mu\nu} $. There are the useful properties
\begin{eqnarray}
\nabla_{\tau} g_{\mu\nu}=0\,,\quad
\nabla_{\tau} e^a_{\mu}=0\,,\quad
\nabla_{\tau} \gamma^{\mu}=0\,.
\label{prop1}
\end{eqnarray}
As a result, we obtain the following expression for the stress-energy tensor in flat space-time
\begin{eqnarray}\label{temres}
T^{\mu\nu}&=& \frac{1}{2} \varepsilon^{\lambda\alpha\beta\rho}\bar{\psi}_{\lambda}\gamma_5 (\gamma^{\mu}\delta^{\nu}_{\alpha}+\gamma^{\nu}\delta^{\mu}_{\alpha})\partial_{\beta}\psi_{\rho} \\
&&+\frac{1}{8}\partial_{\eta}\Big(
\varepsilon^{\lambda\alpha\beta\rho}\bar{\psi}_{\lambda}\gamma_5\gamma_{\alpha} ([\gamma^{\eta},\gamma^{\mu}]\delta^{\nu}_{\beta}+
[\gamma^{\eta},\gamma^{\nu}]\delta^{\mu}_{\beta})\psi_{\rho}\Big)\nonumber \\
&&+\frac{i}{4}\Big(
\bar{\lambda}\gamma^{\nu}\partial^{\mu}\lambda-\partial^{\mu}\bar{\lambda}\gamma^{\nu}\lambda+\bar{\lambda}\gamma^{\mu}\partial^{\nu}\lambda-\partial^{\nu}\bar{\lambda}\gamma^{\mu}\lambda
\Big) \nonumber \\
&&+\frac{i }{2}m\Big(
\bar{\psi}^{\mu}\gamma^{\nu}\lambda-\bar{\lambda}\gamma^{\mu}\psi^{\nu}+\bar{\psi}^{\nu}\gamma^{\mu}\lambda-\bar{\lambda}\gamma^{\nu}\psi^{\mu}
\Big)\,, \nonumber
\end{eqnarray}
where we have omitted the terms equal to zero, when the equations of motion are taken into account. The first two lines correspond to the usual Rarita-Schwinger field theory. When deriving (\ref{temres}) and checking its conservation, it is necessary to use the identity
\begin{eqnarray}\label{eps}
&&\eta^{\vartheta\mu}\varepsilon^{\nu\alpha\beta\gamma}+
\eta^{\vartheta\gamma}\varepsilon^{\mu\nu\alpha\beta}+
\eta^{\vartheta\beta}\varepsilon^{\gamma\mu\nu\alpha}+
\eta^{\vartheta\alpha}\varepsilon^{\beta\gamma\mu\nu}\nonumber \\
&&+
\eta^{\vartheta\nu}\varepsilon^{\alpha\beta\gamma\mu}=0\,. 
\end{eqnarray}
Also the product of three gamma matrices can be expanded
\begin{eqnarray}\label{ggg1}
\gamma^{\mu}\gamma^{\nu}\gamma^{\lambda}=i \varepsilon^{\rho\mu\nu\lambda}\gamma_5\gamma_{\rho}+\eta^{\mu\nu}\gamma^{\lambda}+\eta^{\nu\lambda}\gamma^{\mu}-\eta^{\mu\lambda}\gamma^{\nu}\,.\quad
\end{eqnarray}

Vector and axial currents can be obtained using Noether's theorem for global symmetries $ \psi_{\mu}\to e^{i \alpha+i \beta \gamma_5}\psi_{\mu}$ and $\lambda\to e^{i \alpha+i \beta \gamma_5}\lambda$
\begin{eqnarray}
j^{\mu}_V&=& -i \varepsilon^{\lambda\rho\nu\mu}\bar{\psi}_{\lambda} \gamma_{\nu} \gamma_5 \psi_{\rho} +\bar{\lambda}\gamma^{\mu} \lambda\,, \nonumber \\
j^{\mu}_A&=& -i \varepsilon^{\lambda\rho\nu\mu}\bar{\psi}_{\lambda}\gamma_{\nu}\psi_{\rho} +\bar{\lambda}\gamma^{\mu}\gamma_5 \lambda\,.
\label{currents}
\end{eqnarray}

At the classical level, the conservation laws are satisfied, and the stress-energy tensor is symmetric and traceless
\begin{eqnarray}
&&\partial_{\mu}T^{\mu\nu}=0\,,\quad
\partial_{\mu}j^{\mu}_V=0\,,\quad
\partial_{\mu}j^{\mu}_A=0\,, \nonumber \\
&&T^{\mu}_{\mu}=0\,,\quad
T_{\mu\nu}= T_{\nu\mu}\,.
\label{conserve}
\end{eqnarray}
Note that the trace of the stress-energy tensor turns out to be zero only in the extended theory \cite{Adler:2017shl}, while for the standard Rarita-Schwinger it is non-zero \cite{Das:1978tm}. Before using the equations of motion, we have from (\ref{temres})
\begin{eqnarray}
T^{\mu}_{\mu}&=&
\frac{1}{2}\varepsilon^{\lambda\mu\beta\rho}\Big(\bar{\psi}_{\lambda} \gamma_5 \gamma_{\mu} \partial_{\beta} \psi_{\rho} -
 \partial_{\beta}\bar{\psi}_{\lambda} \gamma_5 \gamma_{\mu} \psi_{\rho} 
\Big) \nonumber \\
&&+i\partial_{\eta}\Big[(\bar{\psi}\gamma)\psi^{\eta}-
\bar{\psi}^{\eta}(\gamma \psi)\Big]\nonumber \\
&&+
\frac{i}{2}\Big[\bar{\lambda}(\gamma\partial)\lambda-(\partial \bar{\lambda} \gamma) \lambda \Big]\nonumber \\
&&+
i m\Big[(\bar{\psi}\gamma)\lambda-\bar{\lambda} (\gamma \psi) \Big]
\,.
\label{tmm1}
\end{eqnarray}
Using now the equations of motion, we would get for the usual Rarita-Schwinger field theory (without any additional fields or ghosts) and for the RSA theory (for arbitrary $ m $)
\begin{eqnarray}
\text{RS theory}:\quad T^{\mu}_{\mu}&=&
i\partial_{\eta}\Big[(\bar{\psi}\gamma)\psi^{\eta}-
\bar{\psi}^{\eta}(\gamma \psi)\Big]\neq 0\,,\nonumber \\
\text{RSA theory}:\quad T^{\mu}_{\mu}&=&
0\,.
\label{tmm2}
\end{eqnarray}
Thus, the dilatation current is conserved only in the extended theory.

As in the case of spin 1/2, a chiral anomaly (\ref{anom}) arises when passing to the quantum field theory, which breaks classical conservation law (\ref{conserve}). However, one can see, that 5 can be obtained as the sum of the nonghost contribution of the Rarita-Schwinger field and the free spin 1/2 field.

Propagators for fields $ \psi_{\mu} $ and $ \lambda $ can be obtained from the path integral by constructing the inverse matrix for the action (\ref{action}). Passing to the Fourier transforms
\begin{eqnarray}
\psi_{\mu}(x)&=&\int \frac{d^4 p}{(2\pi)^4}e^{-ipx}\psi_{\mu}(p)\,, \nonumber \\ 
\lambda(x)&=&\int \frac{d^4 p}{(2\pi)^4}e^{-ipx}\lambda(p)\,,
\label{fur}
\end{eqnarray}
we can write the action in the following matrix form
\begin{eqnarray}
S&=& i \int \frac{d^4 p}{(2\pi)^4} \begin{pmatrix} \bar{\psi}_{\lambda}(p) & \lambda(p) \end{pmatrix}\mathcal{M}\begin{pmatrix}
\psi_{\rho}(p) \\
\lambda(p)
\end{pmatrix},\nonumber \\
\mathcal{M}&=&\begin{pmatrix}
\epsilon^{\lambda \rho \mu \nu}\gamma_5 \gamma_{\mu} p_{\nu} && m \gamma^{\lambda} \\
-m \tilde{\gamma}^{\rho} && -i \slashed{p}
\end{pmatrix}\,.
\label{S_euc}
\end{eqnarray}
The propagators are then defined by the inverse matrix  $ \mathcal{N} $ for which $ \mathcal{M} \mathcal{N} = \begin{pmatrix}
\eta^ {\lambda \sigma} & 0 \\ 0 & 1
\end{pmatrix} $
\begin{eqnarray}
\mathcal{N}&=&\begin{pmatrix}
\frac{i}{2 p^2} \Big[\gamma^{\sigma} \slashed{p} \gamma_{\rho} - 2 \Big(\frac{1}{m^2}+\frac{2}{p^2}\Big) p^{\sigma} p_{\rho}\slashed{p} \Big] && -\frac{\slashed{p}p_{\rho}}{m p^2} \\
\frac{\slashed{p}p^{\sigma}}{m p^2} && 0
\end{pmatrix}\,.\,\,\,\,\,\,\,\,\,\,\,\,\,\,\,\,
\label{N}
\end{eqnarray}
As a result, we have for the time ordered Green functions
\begin{eqnarray}
&&\langle T \psi_a^{\rho}(x) \bar{\psi}_b^{\sigma}(0) \rangle =G_{\psi\bar{\psi}}^{\rho\sigma}(x) 
= \int \frac{d^4 p}{(2 \pi)^4} e^{-ipx} \mathcal{N}_{11} =
\nonumber \\ 
&&=\frac{i}{2(2 \pi)^4}
\int \frac{d^4 p}{p^2}\Bigg(\gamma^{\sigma} \slashed{p} \gamma^{\rho} - 2 \bigg(\frac{1}{m^2}+\frac{2}{p^2}\bigg) p^{\sigma} p^{\rho}\slashed{p} \Bigg)_{ab} 
e^{-i p x}\,, \nonumber \\
&&\langle T \lambda_a(x) \bar{\psi}_b^{\sigma}(0)  \rangle = G_{\lambda\bar{\psi}}^{\sigma}(x)
= \int \frac{d^4 p}{(2 \pi)^4} e^{-ipx} \mathcal{N}_{21}
= \nonumber \\
&&=\frac{1}{(2 \pi)^4}
\int \frac{d^4 p}{ m p^2}\slashed{p}_{ab} p^{\sigma}  e^{-i p x}\,, \nonumber \\
&&\langle T \psi^{\rho}_a(x) \bar{\lambda}_b(0)  \rangle 
= G_{\psi\bar{\lambda}}^{\rho}(x) = \int \frac{d^4 p}{(2 \pi)^4} e^{-ipx} \mathcal{N}_{12}
=\nonumber \\
&&= -\frac{1}{(2 \pi)^4}
\int \frac{d^4 p}{ m p^2}\slashed{p}_{ab} p^{\rho}  e^{-i p x}\,, \nonumber \\
&&\langle T \lambda_a(x) \bar{\lambda}_b(0) \rangle =  G_{\lambda\bar{\lambda}}(x) = \int \frac{d^4 p}{(2 \pi)^4} e^{-ipx} \mathcal{N}_{22}= 0 \,,
\label{prop_p}
\end{eqnarray}
where $ a, b $ are bispinor indices and $ \slashed{p}=p_{\mu}\gamma^{\mu} $. Propagators (\ref{prop_p}) can be written in coordinate representation using the general formula \cite{Grozin:2003ak, Kazakov:2008tr}
\begin{eqnarray}
&&\int\frac{d^{D}p\, e^{ipx}}{p^{2(\lambda+1-\alpha)}} = \frac{i 2^{2 \alpha} \pi^{\lambda+1}  \Gamma(\alpha)}{x^{2 \alpha} \Gamma(\lambda+1-\alpha)}\,,\nonumber \\ 
&&\lambda =1-\varepsilon\,,\quad D=2(\lambda+1)\,,
\label{pik_form}
\end{eqnarray}
in particular, we have for the case when $ p^2 $ is in the denominator
\begin{eqnarray}
\int\frac{d^{4}p\, e^{ipx}}{p^{2}} &=& \frac{4 i \pi^2}{  x^2}\,,\nonumber \\ 
\int\frac{d^{4}p\, e^{ipx} p^{\mu}}{p^{2}} &=& -\frac{8 \pi^2 x^{\mu}}{  x^4}\,,\nonumber \\
\int\frac{d^{4}p\, e^{ipx}p^{\mu}p^{\nu}}{p^{2}} &=& \frac{8 i \pi^2}{  x^4}\Big(\eta^{\mu\nu}-4\frac{x^{\mu}x^{\nu}}{x^2}\Big)\,,\nonumber \\
\int\frac{d^{4}p\, e^{ipx}p^{\mu}p^{\nu}p^{\alpha}}{p^{2}} &=& -\frac{32 \pi^2}{  x^6}\Big(\eta^{\mu\nu}x^{\alpha}+\eta^{\mu\alpha}x^{\nu}+\eta^{\nu\alpha}x^{\mu}\nonumber \\ &&-6\frac{x^{\mu}x^{\nu}x^{\alpha}}{x^2}\Big)\,,
\label{pik_form1}
\end{eqnarray}
and similar formulas for the case with $ p^4 $ in the denominator
\begin{eqnarray}
\int\frac{d^{4}p\, e^{ipx} p^{\mu}}{p^{4}} &=& -\frac{2 \pi^2 x^{\mu}}{  x^2}\,,\nonumber \\
\int\frac{d^{4}p\, e^{ipx}p^{\mu}p^{\nu}}{p^{4}} &=& \frac{2 i \pi^2}{  x^2}\Big(\eta^{\mu\nu}-2\frac{x^{\mu}x^{\nu}}{x^2}\Big)\,,\nonumber \\
\int\frac{d^{4}p\, e^{ipx}p^{\mu}p^{\nu}p^{\alpha}}{p^{4}} &=& -\frac{4 \pi^2}{  x^4}\Big(\eta^{\mu\nu}x^{\alpha}+\eta^{\mu\alpha}x^{\nu}+\eta^{\nu\alpha}x^{\mu}\nonumber \\ &&-4\frac{x^{\mu}x^{\nu}x^{\alpha}}{x^2}\Big)\,. 
\label{pik_form2}
\end{eqnarray}
Using (\ref{pik_form1}) and (\ref{pik_form2}) in (\ref{prop_p}), we obtain the following expressions for the propagators 
\begin{eqnarray}
&&\langle T \psi_a^{\rho}(x) \bar{\psi}_b^{\sigma}(0) \rangle 
=\frac{i}{4 \pi^2 x^4}
\Big[
\gamma^{\sigma}\slashed{x}\gamma^{\rho}-2\Big(1+\frac{4}{m^2 x^2}\Big) \nonumber \\
&&\times (\eta^{\rho\sigma}\slashed{x}+\gamma^{\rho}x^{\sigma}+\gamma^{\sigma}x^{\rho})
+8\left(1+\frac{6}{m^2x^2}\right)\frac{x^{\rho}x^{\sigma}\slashed{x}}{x^2}
\Big]_{ab} \,, \nonumber \\
&&\langle T \lambda_a(x) \bar{\psi}_b^{\sigma}(0)  \rangle 
= \frac{i}{2 \pi^2 m x^4}
\Big(\gamma^{\sigma} - \frac{4 x^{\sigma} \slashed{x}}{x^2}\Big)_{ab}\,, \nonumber \\
&&\langle T \psi^{\rho}_a(x) \bar{\lambda}_b(0)  \rangle 
= \frac{-i}{2 \pi^2 m x^4}
\Big(\gamma^{\rho} - \frac{4 x^{\rho} \slashed{x}}{x^2}\Big)_{ab}\,,
\nonumber \\
&&\langle T \lambda_a(x) \bar{\lambda}_b(0) \rangle = 0 \,.
\label{prop_x}
\end{eqnarray}

Note also that, as was noted in \cite{Adler:2017shl} (see also \cite{Adler:2019zxx}), the ghosts in the RSA theory turn out to be non-propagating and do not contribute to the quantities of interest to us.

%================
\section{Anomaly calculation method: universal form of three-point functions}
\label{sec_method}
%================

There are many ways to find quantum anomalies. In particular, the anomaly (\ref{anom}) in \cite{Adler:2017shl} was derived by the well-known method of shifting in the momentum integrals: before taking the divergence, the three-point function for the mean value of the current allows a shift in the momentum variable under the integral, which at the next stage leads to a finite difference of linearly diverging integrals.

However, it turns out that there is a universal method for calculating anomalies that does not require explicit consideration of divergent integrals. This method is based on an analysis of the form of three-point functions. In \cite{Erdmenger:1996yc}, in this way a gauge chiral anomaly was found, and in \cite{Erdmenger:1999xx} -- a gravitational chiral anomaly. We start with a simpler case with a gauge chiral anomaly.

%================
\subsection{Gauge chiral anomaly}
\label{subsec_method_gauge}
%================

In \cite{Erdmenger:1996yc} it was shown that for conformally symmetric field theories the three-point connected correlator $ \langle T \hat{j}_V^{\mu}(x) \hat{j}_V^{\nu}(y) \hat{j}^{\omega}_{A}(z) \rangle_c $ has a universal form
\begin{eqnarray}
&&\langle T \hat{j}_V^{\mu}(x) \hat{j}_V^{\nu}(y) \hat{j}_A^{\omega}(z) \rangle_c = \nonumber \\
&&= -4 \mathscr{B} \frac{I^{\mu}_{\mu'}(x-z) I^{\nu}_{\nu'}(y-z)}{(x-z)^6 (y-z)^6} \varepsilon^{\mu' \nu' \lambda\omega}\frac{Z_{\lambda}}{Z^4}\,,
\label{jjj_erdm_gen}
\end{eqnarray}
where $ \mathscr{B} $ is an uncertain coefficient, that depends on a particular choice of the field theory. The notations are introduced
\begin{eqnarray}
I_{\mu\nu}(x) & =&  \eta_{\mu \nu} -2 \frac{x_{\mu}x_{\nu}}{x^2}\,,\nonumber \\
Z_{\mu} &=& \frac{(x-z)_{\mu}}{(x-z)^2}-\frac{(x-y)_{\mu}}{(x-y)^2}
\,.
\label{erdm_gauge_anom}
\end{eqnarray}
The Eq. (\ref{jjj_erdm_gen}) was derived using (\ref{conserve}), in particular, the property $ T^{\mu}_{\mu}=0 $.

On the other hand, we can consider the gauge chiral anomaly
\begin{eqnarray}
\langle \partial_{\mu}\hat{j}^{\mu}_A\rangle = -b\frac{\pi^4}{4} \varepsilon^{\mu\nu\alpha\beta}F_{\mu\nu}F_{\alpha\beta}\,.
\label{anom_gauge_gen}
\end{eqnarray}
The central point for us is the relationship between the three-point function and the gauge chiral anomaly
\begin{eqnarray}
\mathscr{B}=b\,.
\label{Bb}
\end{eqnarray}
To substantiate this relationship of the factors, one needs to take the divergence from (\ref{jjj_erdm_gen}) and multiply it by the external fields.

It is technically convenient to consider a special case when all three points in (\ref{jjj_erdm_gen}) lie on the same axis
\begin{eqnarray}
x_{\mu}=x\, e_{\mu}\,,\,
y_{\mu}=y\, e_{\mu}\,,\,
z_{\mu}=z\, e_{\mu}\,,
\label{xyz_simple}
\end{eqnarray}
where $ e^2=\pm 1 $. In this case (\ref{jjj_erdm_gen}) takes the form
\begin{eqnarray}
\langle T \hat{j}_V^{\mu}(x) \hat{j}_V^{\nu}(y) \hat{j}^{\omega}_{A}(z) \rangle =  \frac{4 \mathscr{B} \, e^2 e_{\vartheta } \varepsilon^{\vartheta \mu \nu \omega }}{\pi ^6 (x-y)^3 (x-z)^3 (y-z)^3}\,.\quad
\label{jjj_erdm}
\end{eqnarray}

Thus, to calculate the gauge chiral quantum anomaly, it is enough to find the three-point function $ \langle T \hat{j}_V^{\mu}(x) \hat{j}_V^{\nu}(y) \hat{j}^{\omega}_{A}(z) \rangle $ using the Feynman rules in the coordinate representation. In fact, for this we need to multiply the propagators and find the trace of Dirac matrices. Then it is necessary to check that the three-point function has the form (\ref{jjj_erdm_gen}) or (\ref{jjj_erdm}), and then the factor $ \mathscr{B} $ in front of the universal function determines the coefficient in the gauge chiral anomaly. In particular, this was done for spin 1/2 in \cite{Erdmenger:1996yc}
\begin{eqnarray}
\mathscr{B}_{s=1/2} = \frac{1}{4 \pi^6}\,,
\label{b_dirac}
\end{eqnarray}
which reproduces the standard formula (\ref{anom12}). In Section \ref{subsec_calc_gauge} we will use this method to find the anomaly in the extended spin 3/2 theory.

%================
\subsection{Gravitational chiral anomaly}
\label{subsec_method_grav}
%================

The method described above was generalized in \cite{Erdmenger:1999xx} to the case of a gravitational chiral anomaly.
In this case we have to consider a three-point connected correlator with two stress-energy tensors and one axial current. Then for a conformally symmetric theory it will have the universal form
\begin{eqnarray}
&&\langle T \hat{T}^{\mu\nu}(x) \hat{T}^{\sigma\rho}(y) \hat{j}^{\omega}_{A}(z) \rangle_c = \frac{1}{(x-z)^8(y-z)^8}
\nonumber \\
&&\times \mathscr{I}_T^{\mu\nu,\mu'\nu'}(x-z) \mathscr{I}_T^{\sigma\rho,\sigma'\rho'}(y-z) {t^{TTA}_{\mu'\nu'\sigma'\rho'}}^{\omega}(Z)\,,\qquad
\label{erdm_TTj}
\end{eqnarray}
where the notations are introduced
\begin{eqnarray}\label{erdm_design}
\mathscr{I}^T_{\mu\nu,\sigma\rho}(x) &=&
\mathscr{E}^T_{\mu\nu,\alpha\beta} I^{\alpha}_{\sigma}(x)
I^{\beta}_{\rho}(x)\,, \\
\mathscr{E}^T_{\mu\nu,\alpha\beta} &=& \frac{1}{2}(\eta_{\mu\alpha} \eta_{\nu\beta}+\eta_{\mu\beta} \eta_{\nu\alpha})-\frac{1}{4}\eta_{\mu\nu} \eta_{\alpha\beta}\,,
\nonumber \\
t^{TTA}_{\mu\nu\sigma\rho\omega}(Z) &=& \frac{\mathscr{A}}{Z^6}( {\mathscr{E}^T_{\mu\nu,\eta}}^{\varepsilon} \mathscr{E}^T_{\sigma\rho,\kappa\varepsilon} {\varepsilon_{\omega}}^{\eta\kappa\lambda}Z_{\lambda}\nonumber \\
&& -6\, \mathscr{E}^T_{\mu\nu,\eta\gamma} \mathscr{E}^T_{\sigma\rho,\kappa\delta} {\varepsilon_{\omega}}^{\eta\kappa\lambda} Z^{\gamma}Z^{\delta}Z_{\lambda}Z^{-2})
\,.
\nonumber
\end{eqnarray}
Here we have the same necessary conditions (\ref{conserve}) for (\ref{erdm_TTj}) to be valid, as in the case of a gauge chiral anomaly. In particular, the coefficient 6 in (\ref{erdm_design}) is a direct consequence of the vanishing of the trace of the stress-energy tensor.

On the other hand, it is known that there is a gravitational chiral anomaly in the gravitational field
\begin{eqnarray}
\langle \nabla_{\mu} j^{\mu}_5 \rangle=
a\frac{\pi^4}{384 \sqrt{-g}}\varepsilon^{\mu\nu\rho\sigma}
R_{\mu\nu\kappa\lambda}{R_{\rho\sigma}}^{\kappa\lambda}\,.
\label{anom_erdm}
\end{eqnarray}
Of key importance for us is the equality of the factor $ \mathscr{A} $ to the coefficient $ a $ in the  axial-gravitational anomaly
\begin{eqnarray}
\mathscr{A}=a\,.
\label{Aa}
\end{eqnarray}
%?%
To substantiate this equality, in \cite{Erdmenger:1999xx} it was first shown that despite the fact that the one-point anomalous divergence, like (\ref{anom_erdm}), vanishes in a flat space-time limit, a similar identity for the three-point function (\ref{erdm_TTj}) is not equal to zero. However, the divergence of (\ref{erdm_TTj}) and (\ref{anom_erdm}) are proportional to the same numerical factor. Thus, we can establish the correspondence (\ref{Aa}) and calculate gravitational chiral anomaly from the flat space correlator.

As with the gauge anomaly, calculations can be simplified by assuming that all three points lie on the same axis (\ref{xyz_simple}).
Then (\ref{jjj_erdm_gen}) takes the form
\begin{eqnarray}
&&\langle T\, \hat{T}^{\mu\nu}(x) \hat{T}^{\sigma\rho}(y) \hat{j}^{\omega}_{A}(z) \rangle = \mathscr{A} \Big(4 (x-y)^5 \nonumber \\
&&\times(x-z)^3
(y-z)^3\Big)^{-1}e_{\vartheta } \Big( \eta ^{\nu \rho } \varepsilon^{\sigma \vartheta \mu \omega }+\eta^{\mu \rho }
\varepsilon^{\sigma \vartheta \nu \omega }\nonumber \\
&& +\eta^{\nu \sigma } \varepsilon^{\vartheta \mu \rho \omega }+\eta^{\mu\sigma } \varepsilon^{\vartheta \nu \rho \omega }  -6 e^2 (e^{\nu } e^{\rho } \varepsilon^{\sigma \vartheta \mu\omega }\nonumber \\
&&+e^{\mu } e^{\rho } \varepsilon^{\sigma \vartheta \nu \omega }+e^{\sigma } e^{\nu } \varepsilon^{\vartheta\mu \rho \omega }  +e^{\sigma } e^{\mu } \varepsilon^{\vartheta \nu \rho \omega })\Big)
\,.
\label{ttj_erdm}
\end{eqnarray}

Thus, it is necessary to find a three-point correlator $ \langle \hat{T}(x)\hat{T}(y)\hat{j}_A(z)\rangle $ in an ordinary flat space-time and check that it satisfies (\ref{erdm_TTj}) or (\ref{ttj_erdm}). The coefficient in front of the universal function will give us a factor of the gravitational chiral anomaly. In particular, the anomaly for spin 1/2 was found in this way \cite{Erdmenger:1999xx}
\begin{eqnarray}
\mathscr{A}_{s=1/2} = \frac{1}{\pi^6}\,,
\label{a_dirac}
\end{eqnarray}
which matches (\ref{anom12}). In the next section, we will use the described method to find the gravitational chiral anomaly for the extended theory of spin 3/2.

%================
\section{Calculation of anomalies in the Rarita-Schwinger-Adler theory}
\label{sec_calc}
%================

%================
\subsection{Gauge chiral anomaly}
\label{subsec_calc_gauge}
%================

In contrast to the free Rarita-Schwinger fields, for the extended Rarita-Schwinger-Adler theory (\ref{action}), all of the relations (\ref{conserve}) are satisfied. Vanishing of the trace of the stress-energy tensor allows us to assume the existence of a conformal symmetry, since in most of the known theories with a conserved dilatation current there is also a conformal symmetry \cite{Chernodub:2021nff, Nakayama:2013is}. This allows us to use the method described in the previous section to find the anomalies.

We will start by considering the gauge anomaly for which the result (\ref{anom}) is already known, to demonstrate the features of the methods.

Operators of the currents can be decomposed depending on the field content
\begin{eqnarray}
\hat{j}^{\mu} &=& \hat{j}_{\bar{\psi}\psi}^{\mu}+ \hat{j}_{\bar{\lambda}\lambda}^{\mu}\,, \nonumber \\
\hat{j}_A^{\mu} &=& \hat{j}_{A\bar{\psi}\psi}^{\mu}+ \hat{j}_{A\bar{\lambda}\lambda}^{\mu}\,,
\label{current_fields}
\end{eqnarray}
where
\begin{eqnarray}
\hat{j}_{\bar{\psi}\psi}^{\mu} &=& \mathcal{J}^{\mu\eta_1\eta_2}_{(\bar{\psi}\psi)ab}\bar{\psi}^a_{\eta_1}\psi^b_{\eta_2}\,,\,\,\,\,\, \mathcal{J}^{\mu\eta_1\eta_2}_{(\bar{\psi}\psi)}=-i\varepsilon^{\eta_1\eta_2\nu\mu}\gamma_{\nu}\gamma_5\,, \nonumber \\
\hat{j}_{\bar{\lambda}\lambda}^{\mu}&=&\mathcal{J}^{\mu}_{(\bar{\lambda}\lambda)ab}\bar{\lambda}_a\lambda_b\,,\,\,\,\,\,\,\,\,\,\,\,\,\,\,\,\,
\mathcal{J}^{\mu}_{(\bar{\lambda}\lambda)}=\gamma^{\mu}\,,
\nonumber \\
\hat{j}_{A\bar{\psi}\psi}^{\mu} &=& \mathcal{J}^{\mu\eta_1\eta_2}_{A(\bar{\psi}\psi)ab}\bar{\psi}^a_{\eta_1}\psi^b_{\eta_2}\,,\, 
{\mathcal{J}}^{\mu\eta_1\eta_2}_{A(\bar{\psi}\psi)}=-i\varepsilon^{\eta_1\eta_2\nu\mu}\gamma_{\nu}\,, \nonumber \\
\hat{j}_{A\bar{\lambda}\lambda}^{\mu}&=&\mathcal{J}^{\mu}_{A(\bar{\lambda}\lambda)ab}\bar{\lambda}_a\lambda_b\,,\,\,\,\,\,\,\,\,
{\mathcal{J}}^{\mu}_{A(\bar{\lambda}\lambda)}=\gamma^{\mu}\gamma_5\,.
\label{oper_gauge}
\end{eqnarray}

Following \cite{Adler:2017shl}, we will consider the limit $ m\to \infty $. According to the Wick theorem, we have
\begin{eqnarray}
&&\langle T \bar{\varphi}_1(x) \varphi_2(x) 
\bar{\varphi}_3(y) \varphi_4(y) 
\bar{\varphi}_5(z) \varphi_6(z) \rangle_c = \nonumber \\
&& -\langle T \bar{\varphi}_1(x) \varphi_4(y) \rangle\langle T
 \varphi_2(x)\bar{\varphi}_5(z)  \rangle\langle T
\bar{\varphi}_3(y) \varphi_6(z) \rangle
\nonumber \\
&& +\langle T \bar{\varphi}_1(x) \varphi_6(z) \rangle\langle T
\varphi_2(x)\bar{\varphi}_3(y)  \rangle\langle T
\varphi_4(y) \bar{\varphi}_5(z) \rangle
 \,,\quad
\label{wick}
\end{eqnarray}
where we left only connected contributions. $ \varphi_i $ is either a $ \psi_{\mu} $ field or a $ \lambda $ field.

Then we obtain
\begin{eqnarray}
&&\langle T \hat{j}^{\mu}(x) \hat{j}^{\nu}(y) \hat{j}^{\omega}_{A}(z) \rangle = - \mathrm{tr}\Big\{\mathcal{J}^{\mu\eta_1\eta_2}_{(\bar{\psi}\psi)}G^{\psi\bar{\psi}}_{\eta_2\eta_5}(x-z)\nonumber \\
&&\times\mathcal{J}^{\omega\eta_5\eta_6}_{A(\bar{\psi}\psi)} G^{\psi\bar{\psi}}_{\eta_6\eta_3}(z-y)\mathcal{J}_{(\bar{\psi}\psi)}^{\nu\eta_3\eta_4}G^{\psi\bar{\psi}}_{\eta_4\eta_1}(y-x)
\Big\} \nonumber \\
&& - \mathrm{tr}\Big\{\mathcal{J}_{(\bar{\psi}\psi)}^{\mu\eta_1\eta_2} G^{\psi\bar{\psi}}_{\eta_2\eta_3}(x-y) \mathcal{J}_{(\bar{\psi}\psi)}^{\nu\eta_3\eta_4} G^{\psi\bar{\psi}}_{\eta_4\eta_5}(y-z)\nonumber \\
&&\times \mathcal{J}_{A(\bar{\psi}\psi)}^{\omega\eta_5\eta_6} G^{\psi\bar{\psi}}_{\eta_6\eta_1}(z-x)
\Big\} \,.
\label{corr_gauge}
\end{eqnarray}
All the terms with field $ \lambda $ are to be dropped out: they are either equal to zero due to $ \langle\lambda\bar{\lambda}\rangle=0 $, or tend to zero in the  limit $ m\to \infty $.

The calculations can be simplified by considering the case of codirectional vectors (\ref{xyz_simple}) (the case with arbitrary points is considered in Appendix \ref{append}). Taking into account the limit $ m\to \infty $ the Green's function will be
\begin{eqnarray}
G^{\psi\bar{\psi}}_{\mu\nu}(x)  &\to&
g_{\mu\nu\alpha}(x) \gamma^{\alpha}+f_{\mu\nu\alpha}(x) \gamma_5\gamma^{\alpha}\,, \nonumber \\
g_{\mu\nu\alpha}(x) &=& -\frac{i}{4 \pi^2 x^3}\Big(\eta_{\nu\alpha} e_{\mu}+\eta_{\mu\alpha} e_{\nu} - 8 \frac{e_{\alpha}e_{\mu}e_{\nu}}{e^2}+
3 e_{\alpha} \eta_{\mu\nu}\Big)\,, \nonumber \\
f_{\mu\nu\alpha}(x) &=& -\frac{1}{4 \pi^2 x^3}e^{\beta}\varepsilon_{\mu\nu\alpha\beta}\,,
\label{prop_simple}
\end{eqnarray}
where in $ x^3 $ there is a power of a number, and in $ e^2$ there is a square of a vector. We also used the properties of Dirac matrices, expanding the product of three matrices. Thus, we need to find a trace with only 6 or 7 gamma matrices. Finally, we obtain
\begin{eqnarray}
\langle T \hat{j}^{\mu}(x) \hat{j}^{\nu}(y) \hat{j}^{\omega}_{A}(z) \rangle_c =  \frac{5 e^2 e_{\vartheta } \varepsilon^{\vartheta \mu \nu \omega }}{\pi ^6 (x-y)^3 (x-z)^3 (y-z)^3}\,.\quad\,\,\,
\label{3point_gauge}
\end{eqnarray}
Function (\ref{3point_gauge}) corresponds to the universal conformally symmetric form (\ref{jjj_erdm}) and the coefficient in the anomaly is
\begin{eqnarray}
\mathscr{B}=\frac{5}{4\pi^6}\,.
\label{b_rsa}
\end{eqnarray}
Taking into account (\ref{anom_gauge_gen}) and (\ref{Bb}), we get agreement with the original result (\ref{anom}).

%================
\subsection{Gravitational chiral anomaly}
\label{subsec_calc_grav}
%================

Now we move on to the gravitational chiral anomaly. In this case, we need to find a three-point function with two operators of the stress-energy tensor and one operator of the axial current (\ref{erdm_TTj}). Now the interaction between the fields $ \psi_{\mu} $ and $ \lambda $ plays a significant role, since in the limit $ m\to\infty $ negative powers $ m^{-n} $ from the propagators can be canceled by positive powers $ m^n $ from the operators of the stress-energy tensor in the vertices.

First, we decompose the stress-energy tensor (\ref{temres}) depending on the field content
\begin{eqnarray}
\hat{T}^{\mu\nu}& = & \hat{T}^{\mu\nu}_{\bar{\psi}\psi}+\hat{T}^{\mu\nu}_{\bar{\lambda}\lambda}+\hat{T}^{\mu\nu}_{\bar{\psi}\lambda}+\hat{T}^{\mu\nu}_{\bar{\lambda}\psi}\,,
\label{tensor_fields1}
\end{eqnarray}
where
\begin{eqnarray}
\hat{T}^{\mu\nu}_{\bar{\psi}\psi}&=&\frac{1}{2} \varepsilon^{\lambda\alpha\beta\rho}\bar{\psi}_{\lambda}\gamma_5 (\gamma^{\mu}\delta^{\nu}_{\alpha}+\gamma^{\nu}\delta^{\mu}_{\alpha})\partial_{\beta}\psi_{\rho}\,, \nonumber \\
&&+\frac{1}{8}\partial_{\eta}\Big(
\varepsilon^{\lambda\alpha\beta\rho}\bar{\psi}_{\lambda}\gamma_5\gamma_{\alpha} ([\gamma^{\eta},\gamma^{\mu}]\delta^{\nu}_{\beta}+
[\gamma^{\eta},\gamma^{\nu}]\delta^{\mu}_{\beta})\psi_{\rho}\Big)\,, \nonumber \\
\hat{T}^{\mu\nu}_{\bar{\lambda}\lambda}&=&\frac{i}{4}\Big(
\bar{\lambda}\gamma^{\nu}\partial^{\mu}\lambda-\partial^{\mu}\bar{\lambda}\gamma^{\nu}\lambda+\bar{\lambda}\gamma^{\mu}\partial^{\nu}\lambda-\partial^{\nu}\bar{\lambda}\gamma^{\mu}\lambda
\Big)\,, \nonumber \\
\hat{T}^{\mu\nu}_{\bar{\psi}\lambda} &=& \frac{i }{2}m\Big(
\bar{\psi}^{\mu}\gamma^{\nu}\lambda+\bar{\psi}^{\nu}\gamma^{\mu}\lambda\Big)\,, \nonumber \\
\hat{T}^{\mu\nu}_{\bar{\lambda}\psi} &=& -\frac{i }{2}m\Big(
\bar{\lambda}\gamma^{\mu}\psi^{\nu}+\bar{\lambda}\gamma^{\nu}\psi^{\mu}
\lambda \Big)\,.
\label{tensor_fields2}
\end{eqnarray}
Then the correlator (\ref{erdm_TTj}) splits into 32 terms depending on the set of fields
\begin{eqnarray}
\langle T \hat{T}^{\mu\nu}(x) \hat{T}^{\sigma\rho}(y) \hat{j}^{\omega}_{A}(z) \rangle_c = \langle \hat{T}_{\bar{\psi}\psi} \hat{T}_{\bar{\psi}\psi} \hat{j}^A_{\bar{\psi}\psi} \rangle + (31\, \text{terms}) \,,\quad\,\, \,\, 
\label{32terms}
\end{eqnarray}
where we have omitted some notations for short. A typical diagram is shown in Fig. \ref{fig} on the right.

It is clear from the condition $ \langle\lambda\bar{\lambda}\rangle = 0 $ in (\ref{prop_x}) in advance that 12 terms in (\ref{32terms}) are equal to zero
\begin{eqnarray}\label{terms=0}
&&\langle \hat{T}_{\bar{\psi}\psi}  \hat{T}_{\bar{\lambda}\lambda}  \hat{j}^A_{\bar{\lambda}\lambda} \rangle =
\langle \hat{T}_{\bar{\psi}\lambda}  \hat{T}_{\bar{\psi}\lambda}  \hat{j}^A_{\bar{\lambda}\lambda} \rangle =
\langle \hat{T}_{\bar{\psi}\lambda}  \hat{T}_{\bar{\lambda}\psi}  \hat{j}^A_{\bar{\lambda}\lambda} \rangle=  \\
&& =
\langle \hat{T}_{\bar{\psi}\lambda}  \hat{T}_{\bar{\lambda}\lambda}  \hat{j}^A_{\bar{\lambda}\lambda} \rangle =
\langle \hat{T}_{\bar{\lambda}\psi}  \hat{T}_{\bar{\psi}\lambda}  \hat{j}^A_{\bar{\lambda}\lambda} \rangle =
\langle \hat{T}_{\bar{\lambda}\psi}  \hat{T}_{\bar{\lambda}\psi}  \hat{j}^A_{\bar{\lambda}\lambda} \rangle =  \nonumber \\
&&=
\langle \hat{T}_{\bar{\lambda}\psi}  \hat{T}_{\bar{\lambda}\lambda}  \hat{j}^A_{\bar{\lambda}\lambda} \rangle =
\langle \hat{T}_{\bar{\lambda}\lambda}  \hat{T}_{\bar{\psi}\psi}  \hat{j}^A_{\bar{\lambda}\lambda} \rangle =
\langle \hat{T}_{\bar{\lambda}\lambda}  \hat{T}_{\bar{\psi}\lambda}  \hat{j}^A_{\bar{\lambda}\lambda} \rangle  = \nonumber \\
&&=
\langle \hat{T}_{\bar{\lambda}\lambda}  \hat{T}_{\bar{\lambda}\psi}  \hat{j}^A_{\bar{\lambda}\lambda} \rangle =
\langle \hat{T}_{\bar{\lambda}\lambda}  \hat{T}_{\bar{\lambda}\psi}  \hat{j}^A_{\bar{\psi}\psi} \rangle =\langle \hat{T}_{\bar{\lambda}\lambda}  \hat{T}_{\bar{\lambda}\lambda}  \hat{j}^A_{\bar{\lambda}\lambda} \rangle =0\,,\qquad \nonumber
\end{eqnarray}
since they contain one or more propagators with two fields $ \lambda $.
Also, in the $ m\to \infty $ limit, 11 more terms vanish
\begin{eqnarray}
&&\langle \hat{T}_{\bar{\psi}\psi}  \hat{T}_{\bar{\psi}\psi}  \hat{j}^A_{\bar{\lambda}\lambda} \rangle, 
\langle \hat{T}_{\bar{\psi}\psi}  \hat{T}_{\bar{\psi}\lambda}  \hat{j}^A_{\bar{\lambda}\lambda} \rangle,
\langle \hat{T}_{\bar{\psi}\psi}  \hat{T}_{\bar{\lambda}\psi}  \hat{j}^A_{\bar{\lambda}\lambda} \rangle, \nonumber \\
&&
\langle \hat{T}_{\bar{\psi}\psi}  \hat{T}_{\bar{\lambda}\lambda}  \hat{j}^A_{\bar{\psi}\psi} \rangle, 
\langle \hat{T}_{\bar{\psi}\lambda}  \hat{T}_{\bar{\psi}\psi}  \hat{j}^A_{\bar{\lambda}\lambda} \rangle,
\langle \hat{T}_{\bar{\psi}\lambda}  \hat{T}_{\bar{\lambda}\lambda}  \hat{j}^A_{\bar{\psi}\psi} \rangle,\nonumber \\
&&
\langle \hat{T}_{\bar{\lambda}\psi}  \hat{T}_{\bar{\psi}\psi}  \hat{j}^A_{\bar{\lambda}\lambda} \rangle,
\langle \hat{T}_{\bar{\lambda}\psi}  \hat{T}_{\bar{\lambda}\lambda}  \hat{j}^A_{\bar{\psi}\psi} \rangle,
\langle \hat{T}_{\bar{\lambda}\lambda}  \hat{T}_{\bar{\psi}\psi}  \hat{j}^A_{\bar{\psi}\psi} \rangle,\nonumber \\
&&
\langle \hat{T}_{\bar{\lambda}\lambda}  \hat{T}_{\bar{\psi}\lambda}  \hat{j}^A_{\bar{\psi}\psi} \rangle,
\langle \hat{T}_{\bar{\lambda}\lambda}  \hat{T}_{\bar{\lambda}\psi}  \hat{j}^A_{\bar{\psi}\psi} \rangle \to 0\quad (m \to \infty)
\,.\qquad
\label{terms_to_0}
\end{eqnarray}
Since negative powers from the propagators $ \langle\psi\bar{\lambda}\rangle, \langle\lambda\bar{\psi}\rangle\sim 1/m$ are not compensated by positive powers from the operators $ \hat{T}^{\mu\nu}_{\bar{\psi}\lambda},\hat{T}^{\mu\nu}_{\bar{\lambda}\psi}\sim m $. Thus, in contrast to the gauge anomaly, the terms of interaction with the $ \lambda $ field start to play a role. Finally, in (\ref{32terms}), only 9 terms remain
\begin{eqnarray}
&&\langle T \hat{T}_{\mu\nu}(x) \hat{T}_{\sigma\rho}(y) \hat{j}_{\omega}^{A}(z) \rangle_c =
\langle \hat{T}_{\bar{\psi}\psi} \hat{T}_{\bar{\psi}\psi} \hat{j}^A_{\bar{\psi}\psi} \rangle \nonumber \\
&& +
\langle \hat{T}_{\bar{\psi}\psi} \hat{T}_{\bar{\psi}\lambda} \hat{j}^A_{\bar{\psi}\psi} \rangle +
\langle \hat{T}_{\bar{\psi}\psi} \hat{T}_{\bar{\lambda}\psi} \hat{j}^A_{\bar{\psi}\psi} \rangle +
\langle \hat{T}_{\bar{\psi}\lambda} \hat{T}_{\bar{\psi}\psi} \hat{j}^A_{\bar{\psi}\psi} \rangle \nonumber \\
&&+
\langle \hat{T}_{\bar{\psi}\lambda} \hat{T}_{\bar{\psi}\lambda} \hat{j}^A_{\bar{\psi}\psi} \rangle +
\langle \hat{T}_{\bar{\psi}\lambda} \hat{T}_{\bar{\lambda}\psi} \hat{j}^A_{\bar{\psi}\psi} \rangle+
\langle \hat{T}_{\bar{\lambda}\psi} \hat{T}_{\bar{\psi}\psi} \hat{j}^A_{\bar{\psi}\psi} \rangle \nonumber \\
&&+
\langle \hat{T}_{\bar{\lambda}\psi} \hat{T}_{\bar{\psi}\lambda} \hat{j}^A_{\bar{\psi}\psi} \rangle  +
\langle \hat{T}_{\bar{\lambda}\psi} \hat{T}_{\bar{\lambda}\psi} \hat{j}^A_{\bar{\psi}\psi} \rangle 
\,,
\label{9terms}
\end{eqnarray}
where all the terms except the first one give a finite contribution just due to the cancellation of the powers of $ m $. Note that the terms of the order of $ 1/m^2 $ in the propagator $ \langle\psi\bar{\psi}\rangle$ can obviously be omitted in advance.

Correlators from (\ref{9terms}) can be calculated using the Feynman rules. It is convenient to represent all the operators in a split form
\begin{eqnarray} \nonumber 
\hat{T}^{\sigma\tau}_{\bar{\psi}\psi}(x) &=& \lim_{x_1,x_2\to x} \mathcal{D}^{\sigma\tau\eta\xi}_{(\bar{\psi}\psi)a b}(\partial_{x_1},\partial_{x_2})\bar{\psi}_{\eta a}(x_1)\psi_{\xi b}(x_2)\,,\\
\hat{T}^{\sigma\tau}_{\bar{\psi}\lambda}(x) &=& \lim_{x_1,x_2\to x} \mathcal{D}^{\sigma\tau\eta}_{(\bar{\psi}\lambda)a b} \bar{\psi}_{\eta a}(x_1)\lambda_{b}(x_2)\,, \nonumber\\
\hat{T}^{\sigma\tau}_{\bar{\lambda}\psi}(x) &=& \lim_{x_1,x_2\to x}   \mathcal{D}^{\sigma\tau\eta}_{(\bar{\lambda}\psi)a b} \bar{\lambda}_{ a}(x_1)\psi_{\eta b}(x_2)\,, \nonumber\\
\hat{j}^{\sigma}_{A\bar{\psi}\psi}(x) &=& \lim_{x_1,x_2\to x}  \mathcal{J}^{\sigma \eta \xi}_{A(\bar{\psi}\psi)a b}\bar{\psi}_{\eta a}(x_1)\psi_{\xi b}(x_2)\,,\nonumber \\
\mathcal{D}^{\sigma\tau\eta\xi}_{(\bar{\psi}\psi)}(\partial_{x_1},\partial_{x_2})&=& 
\frac{1}{2} \varepsilon^{\eta \xi \alpha\beta}\gamma_5
(\gamma^{\sigma}\delta^{\tau}_{\alpha}+
\gamma^{\tau}\delta^{\sigma}_{\alpha})
\partial_{\beta}^{x_2} \nonumber \\
&&+\frac{1}{8}
\varepsilon^{\eta \xi \alpha\beta}\gamma_5 \gamma_{\alpha}([\gamma^{\vartheta},\gamma^{\sigma}]\delta^{\tau}_{\beta} \nonumber \\
&&+[\gamma^{\vartheta},\gamma^{\tau}]\delta^{\sigma}_{\beta}) (\partial^{x_1}_{\vartheta}+\partial^{x_2}_{\vartheta})\,, \nonumber \\
\mathcal{D}^{\sigma\tau\eta}_{(\bar{\psi}\lambda)}&=&\frac{i}{2}m 
(\gamma^{\sigma}\eta^{\tau \eta} + \gamma^{\tau}\eta^{\sigma \eta}
)\,, \nonumber \\
\mathcal{D}^{\sigma\tau\eta}_{(\bar{\lambda}\psi)}&=& -\frac{i}{2}m 
(\gamma^{\sigma}\eta^{\tau \eta} + \gamma^{\tau}\eta^{\sigma \eta}
) \,,
\label{split}
\end{eqnarray}
and $ \mathcal{J}^{\sigma \eta \xi}_{A(\bar{\psi}\psi)} $ was written out earlier in (\ref{oper_gauge}). Each of the 9 terms in (\ref{9terms}) splits into two, due to Wick's theorem (\ref{wick}). In particular, the first term in (\ref{9terms}) has the form
\begin{widetext}
\begin{eqnarray}
\langle T \hat{T}^{\mu\nu}_{\bar{\psi}\psi}(x) \hat{T}^{\sigma\rho}_{\bar{\psi}\psi}(y) \hat{j}^{\omega}_{A \bar{\psi}\psi}(z) \rangle_c  &=&
\lim_{
	\def\arraystretch{0.5}\begin{array}{ll}
	{\scriptscriptstyle x_1,x_2\to x}\vspace{0.1mm}\\
	{\scriptscriptstyle y_1,y_2\to y}\vspace{0.1mm}\\
	{\scriptscriptstyle z_1,z_2\to z}
	\end{array}}\bigg(
- \mathrm{tr}\Big\{\mathcal{D}^{\mu\nu\eta_1\eta_2}_{(\bar{\psi}\psi)}
(\partial^{x_1},\partial^{x_2})
G^{\psi\bar{\psi}}_{\eta_2\eta_5}(x_2-z_1)
\mathcal{J}^{\omega\eta_5\eta_6}_{A(\bar{\psi}\psi)} G^{\psi\bar{\psi}}_{\eta_6\eta_3}(z_2-y_1)\nonumber \\
&& \times\mathcal{D}_{(\bar{\psi}\psi)}^{\sigma\rho\eta_3\eta_4}
(\partial^{y_1},\partial^{y_2})
G^{\psi\bar{\psi}}_{\eta_4\eta_1}(y_2-x_1)
\Big\} - \mathrm{tr}\Big\{\mathcal{D}_{(\bar{\psi}\psi)}^{\mu\nu\eta_1\eta_2}
(\partial^{x_1},\partial^{x_2}) G^{\psi\bar{\psi}}_{\eta_2\eta_3}(x_2-y_1) \nonumber \\
&& \times\mathcal{D}_{(\bar{\psi}\psi)}^{\sigma\rho\eta_3\eta_4}
(\partial^{y_1},\partial^{y_2}) G^{\psi\bar{\psi}}_{\eta_4\eta_5}(y_2-z_1)\mathcal{J}_{A(\bar{\psi}\psi)}^{\omega\eta_5\eta_6} G^{\psi\bar{\psi}}_{\eta_6\eta_1}(z_2-x_1)
\Big\} \bigg) \,,
\label{corr_example1}
\end{eqnarray}
where the derivatives act on the function regardless of whether it is on the left or on the right. The rest of the terms in (\ref{9terms}) contain $\lambda$ fields, either one or two. In particular, we obtain
\begin{eqnarray}
\langle T \hat{T}^{\mu\nu}_{\bar{\psi}\psi}(x) \hat{T}^{\sigma\rho}_{\bar{\psi}\lambda}(y) \hat{j}^{\omega}_{A \bar{\psi}\psi}(z) \rangle_c  &=&
\lim_{
	\def\arraystretch{0.5}\begin{array}{ll}
	{\scriptscriptstyle x_1,x_2\to x}\vspace{0.1mm}\\
	{\scriptscriptstyle y_1,y_2\to y}\vspace{0.1mm}\\
	{\scriptscriptstyle z_1,z_2\to z}
	\end{array}}\bigg(
- \mathrm{tr}\Big\{\mathcal{D}^{\mu\nu\eta_1\eta_2}_{(\bar{\psi}\psi)}
(\partial^{x_1},\partial^{x_2})
G^{\psi\bar{\psi}}_{\eta_2\eta_4}(x_2-z_1)
\mathcal{J}^{\omega\eta_4\eta_5}_{A(\bar{\psi}\psi)} G^{\psi\bar{\psi}}_{\eta_5\eta_3}(z_2-y_1)   \nonumber \\
&&\times\mathcal{D}_{(\bar{\psi}\lambda)}^{\sigma\rho\eta_3}
G^{\lambda\bar{\psi}}_{\eta_1}(y_2-x_1)
\Big\}- \mathrm{tr}\Big\{\mathcal{D}_{(\bar{\psi}\psi)}^{\mu\nu\eta_1\eta_2}
(\partial^{x_1},\partial^{x_2}) G^{\psi\bar{\psi}}_{\eta_2\eta_3}(x_2-y_1) \mathcal{D}_{(\bar{\psi}\lambda)}^{\sigma\rho\eta_3}
 \nonumber \\
&&\times G^{\lambda\bar{\psi}}_{\eta_4}(y_2-z_1)\mathcal{J}_{A(\bar{\psi}\psi)}^{\omega\eta_4\eta_5} G^{\psi\bar{\psi}}_{\eta_5\eta_1}(z_2-x_1)
\Big\} \bigg) \,,
\label{corr_example3}
\end{eqnarray}
\begin{eqnarray}
\langle T \hat{T}^{\mu\nu}_{\bar{\psi}\lambda}(x) \hat{T}^{\sigma\rho}_{\bar{\psi}\lambda}(y) \hat{j}^{\omega}_{A\bar{\psi}\psi}(z) \rangle_c  &=&
\lim_{
	\def\arraystretch{0.5}\begin{array}{ll}
	{\scriptscriptstyle x_1,x_2\to x}\vspace{0.1mm}\\
	{\scriptscriptstyle y_1,y_2\to y}\vspace{0.1mm}\\
	{\scriptscriptstyle z_1,z_2\to z}
	\end{array}}\bigg(
- \mathrm{tr}\Big\{\mathcal{D}^{\mu\nu\eta_1}_{(\bar{\psi}\lambda)}
G^{\lambda\bar{\psi}}_{\eta_3}(x_2-z_1)
\mathcal{J}^{\omega\eta_3\eta_4}_{A(\bar{\psi}\psi)} G^{\psi\bar{\psi}}_{\eta_4\eta_2}(z_2-y_1) \mathcal{D}_{(\bar{\psi}\lambda)}^{\sigma\rho\eta_2}
G^{\lambda\bar{\psi}}_{\eta_1}(y_2-x_1)
\Big\} \nonumber \\
&&- \mathrm{tr}\Big\{\mathcal{D}_{(\bar{\psi}\lambda)}^{\mu\nu\eta_1}
G^{\lambda\bar{\psi}}_{\eta_2}(x_2-y_1)  \mathcal{D}_{(\bar{\psi}\lambda)}^{\sigma\rho\eta_2}
G^{\lambda\bar{\psi}}_{\eta_3}(y_2-z_1)\mathcal{J}_{A(\bar{\psi}\psi)}^{\omega\eta_3\eta_4} G^{\psi\bar{\psi}}_{\eta_4\eta_1}(z_2-x_1)
\Big\} \bigg) \,,
\label{corr_example2}
\end{eqnarray}
\begin{eqnarray}
\langle T \hat{T}^{\mu\nu}_{\bar{\psi}\lambda}(x) \hat{T}^{\sigma\rho}_{\bar{\lambda}\psi}(y) \hat{j}^{\omega}_{A\bar{\psi}\psi}(z) \rangle_c  &=&
-	\lim_{
	\def\arraystretch{0.5}\begin{array}{ll}
	{\scriptscriptstyle x_1,x_2\to x}\vspace{0.1mm}\\
	{\scriptscriptstyle y_1,y_2\to y}\vspace{0.1mm}\\
	{\scriptscriptstyle z_1,z_2\to z}
	\end{array}}
\mathrm{tr}\Big\{\mathcal{D}^{\mu\nu\eta_1}_{(\bar{\psi}\lambda)}
G^{\lambda\bar{\psi}}_{\eta_3}(x_2-z_1)
\mathcal{J}^{\omega\eta_3\eta_4}_{A(\bar{\psi}\psi)} G^{\psi\bar{\lambda}}_{\eta_4}(z_2-y_1) \mathcal{D}_{(\bar{\lambda}\psi)}^{\sigma\rho\eta_2}
G^{\psi\bar{\psi}}_{\eta_2\eta_1}(y_2-x_1)
\Big\}  \,.
\label{corr_example4}
\end{eqnarray}
\end{widetext}
The equation (\ref{corr_example4}) contains only one term due to $ \langle\lambda\bar{\lambda}\rangle = 0 $. 

Now we need to use the general form of the propagators (\ref{prop_x}) and take the derivatives coming from the operators of the stress-energy tensor and after that we can use (\ref{xyz_simple}). To simplify calculations, it is convenient to expand the products with three Dirac matrices, coming from (\ref{prop_x}) and (\ref{split}). This will reduce the number of Dirac matrices under the trace from 16 to 6 or 7. So, for $ \mathcal{D}^{\sigma\tau\eta\xi}_{(\bar{\psi}\psi)}(\partial_{x_1},\partial_{x_2}) $ using (\ref{ggg1}) and (\ref{xyz_simple}), we get
\begin{eqnarray}
\mathcal{D}^{\sigma\tau\eta\xi}_{(\bar{\psi}\psi)}(\partial_{x_1},\partial_{x_2})&=&(\partial^{x_1}+\partial^{x_2})_{\vartheta}\gamma_{\mu}A_1^{\sigma\tau\eta\xi\mu\vartheta}\nonumber \\ &&+(\partial^{x_1}_{\vartheta}-\partial^{x_2}_{\vartheta})\gamma_5\gamma_{\mu}A_2^{\sigma\tau\eta\xi\mu\vartheta}\,,\nonumber \\
A_1^{\sigma\tau\eta\xi\mu\vartheta} &=&
\frac{i}{4} (2 \eta ^{\eta \mu } \eta ^{\sigma \tau } \eta ^{\vartheta \xi }-\eta ^{\tau \eta } \eta ^{\sigma \mu }
\eta ^{\vartheta \xi }-\eta ^{\sigma \eta } \eta ^{\tau \mu } \eta ^{\vartheta \xi }\nonumber \\ && -\eta^{\eta \mu } \eta ^{\sigma \xi }
\eta ^{\vartheta \tau }-2 \eta ^{\vartheta \eta } \eta ^{\xi \mu } \eta ^{\sigma \tau }+\eta ^{\sigma \eta } \eta ^{\xi \mu
} \eta ^{\vartheta \tau }\nonumber \\ &&+\eta ^{\tau \eta } \eta ^{\xi \mu } \eta ^{\vartheta \sigma }+\eta ^{\vartheta \eta } \eta ^{\tau
	\mu } \eta ^{\sigma \xi }-\eta ^{\eta \mu } \eta ^{\tau \xi } \eta ^{\vartheta \sigma }\nonumber \\ &&+\eta ^{\vartheta \eta } \eta
^{\sigma \mu } \eta ^{\tau \xi })\,,
\nonumber \\
A_2^{\sigma\tau\eta\xi\mu\vartheta} &=&
\frac{1}{4} \left(\eta ^{\mu \tau } \varepsilon ^{\eta \xi  \vartheta \sigma }+\eta ^{\mu \sigma } \varepsilon ^{\eta \xi  \vartheta \tau }\right)\,.
\label{Dpsps3}
\end{eqnarray}
The propagator $ G_{\psi\bar{\lambda}}^{\mu} $ will take the form
\begin{eqnarray}
G_{\psi\bar{\lambda}}^{\mu}(x)\to\frac{i \gamma_{\eta}}{2 \pi ^2 m\, x^4}\left(4 e^2 e^{\eta } e^{\mu }-\eta ^{\eta \mu }\right)\,,
\label{prop_simple1}
\end{eqnarray}
and the propagator $ G_{\psi\bar{\psi}}^{\mu\nu} $ was given in (\ref{prop_simple}). 

Similarly, derivatives of the propagators can be simplified
\begin{eqnarray}
\partial^{\alpha} G_{\psi\bar{\psi}}^{\mu\nu}(x)&\to& 
-\frac{i \gamma_{\eta}}{4 \pi ^2 x^4}\Big\{\eta ^{\eta \mu } \eta ^{\alpha \nu }+\eta ^{\eta \nu } \eta ^{\alpha \mu }+3 \eta ^{\eta \alpha } \eta ^{\mu
	\nu } \nonumber \\ &&-4 e^2 e^{\alpha } e^{\nu } \eta ^{\eta \mu }-4 e^2 e^{\mu } \left(2 e^{\nu } \eta ^{\eta \alpha }+e^{\alpha } \eta
^{\eta \nu }\right)\nonumber \\ &&+4 e^{\eta } \big[3 e^{\alpha } (4 e^{\mu } e^{\nu }-e^2 \eta ^{\mu \nu })-2 e^2
(e^{\nu } \eta ^{\alpha \mu }\nonumber \\ &&+e^{\mu } \eta ^{\alpha \nu })\big]\Big\}
-\frac{\gamma_5 \gamma_{\eta} \varepsilon ^{\sigma \eta \mu \nu }}{4 \pi ^2 x^4}  \left(4 e^2 e^{\alpha } e_{\sigma }-\delta _{\sigma}^{\alpha }\right)\,,
\nonumber \\
\partial^{\alpha} G_{\psi\bar{\lambda}}^{\mu}(x)&\to& \frac{2 i }{\pi ^2 m\, x^5} \big[e^2  (e^{\mu } \eta ^{\eta \alpha }+e^{\alpha } \eta ^{\eta \mu } )\nonumber \\ &&
+e^{\eta } (e^2 \eta ^{\alpha
	\mu }-6 e^{\alpha } e^{\mu })\big]\,.
\label{dprop_simple}
\end{eqnarray}
Also second order derivatives can be obtained from (\ref{prop_x}) using (\ref{xyz_simple}). The formulas for $ G_{\lambda\bar{\psi}}^{\mu}$ differ only in common sign from $ G_{\psi\bar{\lambda}}^{\mu}$.

Now we can calculate (\ref{corr_example1})-(\ref{corr_example4}), using (\ref{Dpsps3}), (\ref{split}), (\ref{prop_simple}), (\ref{prop_simple1}) and (\ref{dprop_simple}). The most time consuming is to find the term $ \langle \hat{T}_{\bar{\psi}\psi} \hat{T}_{\bar{\psi}\psi} \hat{j}^A_{\bar{\psi}\psi} \rangle $, since it contains the largest number of derivatives and the longest Green's function. Wherein some of the terms are equal to each other up to a change of variables. Namely
\begin{eqnarray}
&&\langle T\, \hat{T}_{\bar{\psi}\psi}^{\mu\nu}(x) \hat{T}_{\bar{\psi}\lambda}^{\sigma\rho}(y) \hat{j}_{A\bar{\psi}\psi}^{\omega}(z) \rangle_c= \nonumber \\
&&=
\langle T\, \hat{T}_{\bar{\psi}\psi}^{\mu\nu}(x) \hat{T}_{\bar{\lambda}\psi}^{\sigma\rho}(y) \hat{j}_{A\bar{\psi}\psi}^{\omega}(z) \rangle_c= \nonumber \\
&&=
-\langle T\, \hat{T}_{\bar{\psi}\lambda}^{\mu\nu}(y) \hat{T}_{\bar{\psi}\psi}^{\sigma\rho}(x) \hat{j}_{A\bar{\psi}\psi}^{\omega}(z) \rangle_c= \nonumber \\
&&=
-\langle T\, \hat{T}_{\bar{\lambda}\psi}^{\mu\nu}(y) \hat{T}_{\bar{\psi}\psi}^{\sigma\rho}(x) \hat{j}_{A\bar{\psi}\psi}^{\omega}(z) \rangle_c\,,
\label{equal1}
\end{eqnarray}
also, one could check, that
\begin{eqnarray}
&&\langle T\, \hat{T}_{\bar{\lambda}\psi}^{\mu\nu}(x) \hat{T}_{\bar{\psi}\lambda}^{\sigma\rho}(y) \hat{j}_{A\bar{\psi}\psi}^{\omega}(z) \rangle_c =\nonumber \\
&& =
\langle T\, \hat{T}_{\bar{\psi}\lambda}^{\mu\nu}(x) \hat{T}_{\bar{\lambda}\psi}^{\sigma\rho}(y) \hat{j}_{A\bar{\psi}\psi}^{\omega}(z) \rangle_c\,,
\label{equal2}
\end{eqnarray}
and
\begin{eqnarray}
&&\langle T\, \hat{T}_{\bar{\lambda}\psi}^{\mu\nu}(x) \hat{T}_{\bar{\lambda}\psi}^{\sigma\rho}(y) \hat{j}_{A\bar{\psi}\psi}^{\omega}(z) \rangle_c =\nonumber \\
&&=
\langle T\, \hat{T}_{\bar{\psi}\lambda}^{\mu\nu}(x) \hat{T}_{\bar{\psi}\lambda}^{\sigma\rho}(y) \hat{j}_{A\bar{\psi}\psi}^{\omega}(z) \rangle_c\,.
\label{equal3}
\end{eqnarray}
Therefore, only 4 matrix elements are independent in (\ref{9terms}). Finally we obtain
\begin{widetext}
\begin{eqnarray}\label{1}
&&\langle T\, \hat{T}_{\bar{\psi}\psi}^{\mu\nu}(x) \hat{T}_{\bar{\psi}\psi}^{\sigma\rho}(y) \hat{j}_{A\bar{\psi}\psi}^{\omega}(z) \rangle_c = \frac{1}{4 \pi ^6 (x-y)^5 (x-z)^4 (y-z)^4} e_{\vartheta } (28 e^2 x^2 e^{\mu } e^{\sigma } \varepsilon ^{\vartheta \nu \rho \omega
}-\varepsilon ^{\sigma \vartheta \nu \omega } (2 e^2 e^{\mu } e^{\rho } (-14 x^2\nonumber \\
&&+9 z(x+y)+19 x y-14 y^2-9 z^2)+(26 x^2-3 z (x+y)-49 x y+26 y^2+3 z^2) \eta ^{\mu \rho })+2
e^2 e^{\nu } (14 x^2-19 x y\nonumber \\
&&-9 x z+14 y^2-9 y z+9 z^2) (e^{\rho } \varepsilon ^{\sigma
	\vartheta \mu \omega }+e^{\sigma } \varepsilon ^{\vartheta \mu \rho \omega })-38 e^2 x y e^{\mu
} e^{\sigma } \varepsilon ^{\vartheta \nu \rho \omega }-18 e^2 x z e^{\mu } e^{\sigma } \varepsilon
^{\vartheta \nu \rho \omega }+28 e^2 y^2 e^{\mu } e^{\sigma } \varepsilon ^{\vartheta \nu \rho \omega
}\nonumber \\
&&-18 e^2 y z e^{\mu } e^{\sigma } \varepsilon ^{\vartheta \nu \rho \omega }+18 e^2 z^2 e^{\mu
} e^{\sigma } \varepsilon ^{\vartheta \nu \rho \omega }-26 x^2 \eta ^{\nu \rho } \varepsilon ^{\sigma
	\vartheta \mu \omega }-26 x^2 \eta ^{\mu \sigma } \varepsilon ^{\vartheta \nu \rho \omega }-(26 x^2-3 z
(x+y)-49 x y\nonumber \\
&&+26 y^2+3 z^2) \eta ^{\nu \sigma } \varepsilon ^{\vartheta \mu \rho \omega }+49 x y \eta
^{\nu \rho } \varepsilon ^{\sigma \vartheta \mu \omega }+49 x y \eta ^{\mu \sigma } \varepsilon ^{\vartheta
	\nu \rho \omega }+3 x z \eta ^{\nu \rho } \varepsilon ^{\sigma \vartheta \mu \omega }+3 x z \eta ^{\mu \sigma
} \varepsilon ^{\vartheta \nu \rho \omega }-26 y^2 \eta ^{\nu \rho } \varepsilon ^{\sigma \vartheta \mu
	\omega }\nonumber \\
&&-26 y^2 \eta ^{\mu \sigma } \varepsilon ^{\vartheta \nu \rho \omega }+3 y z \eta ^{\nu \rho }
\varepsilon ^{\sigma \vartheta \mu \omega }+3 y z \eta ^{\mu \sigma } \varepsilon ^{\vartheta \nu \rho \omega
}-3 z^2 \eta ^{\nu \rho } \varepsilon ^{\sigma \vartheta \mu \omega }-3 z^2 \eta ^{\mu \sigma } \varepsilon
^{\vartheta \nu \rho \omega })\,,
\end{eqnarray}
\begin{eqnarray}
&&\langle T\, \hat{T}_{\bar{\psi}\psi}^{\mu\nu}(x) \hat{T}_{\bar{\psi}\lambda}^{\sigma\rho}(y) \hat{j}_{A\bar{\psi}\psi}^{\omega}(z) \rangle_c = \frac{1}{4 \pi ^6 (x-y)^5 (x-z)^4 (y-z)^4}e_{\vartheta } (4 \text{e}^2 x^2 e^{\mu } e^{\sigma } \varepsilon ^{\vartheta \nu \rho \omega
}-\varepsilon ^{\sigma \vartheta \nu \omega } (2 \text{e}^2 e^{\mu } e^{\rho } (-2 x^2\nonumber \\
&&+3 z (7
x+y)-17 x y+7 y^2-12 z^2)+(10 x^2+7 x y-27 x z-13 y^2+19 y z+4 z^2) \eta ^{\mu \rho
})+2 \text{e}^2 e^{\nu } (2 x^2\nonumber \\
&&+17 x y-21 x z-7 y^2-3 y z+12 z^2) (e^{\rho }
\varepsilon ^{\sigma \vartheta \mu \omega }+e^{\sigma } \varepsilon ^{\vartheta \mu \rho \omega })+34
\text{e}^2 x y e^{\mu } e^{\sigma } \varepsilon ^{\vartheta \nu \rho \omega }-42 \text{e}^2 x z e^{\mu }
e^{\sigma } \varepsilon ^{\vartheta \nu \rho \omega }\nonumber \\
&&-14 \text{e}^2 y^2 e^{\mu } e^{\sigma } \varepsilon
^{\vartheta \nu \rho \omega }-6 \text{e}^2 y z e^{\mu } e^{\sigma } \varepsilon ^{\vartheta \nu \rho \omega
}+24 \text{e}^2 z^2 e^{\mu } e^{\sigma } \varepsilon ^{\vartheta \nu \rho \omega }-10 x^2 \eta ^{\nu \rho }
\varepsilon ^{\sigma \vartheta \mu \omega }-10 x^2 \eta ^{\mu \sigma } \varepsilon ^{\vartheta \nu \rho
	\omega }-(10 x^2+7 x y\nonumber \\
&&-27 x z-13 y^2+19 y z+4 z^2) \eta ^{\nu \sigma } \varepsilon ^{\vartheta \mu
	\rho \omega }-7 x y \eta ^{\nu \rho } \varepsilon ^{\sigma \vartheta \mu \omega }-7 x y \eta ^{\mu \sigma }
\varepsilon ^{\vartheta \nu \rho \omega }+27 x z \eta ^{\nu \rho } \varepsilon ^{\sigma \vartheta \mu \omega
}+27 x z \eta ^{\mu \sigma } \varepsilon ^{\vartheta \nu \rho \omega }\nonumber \\
&&+13 y^2 \eta ^{\nu \rho } \varepsilon
^{\sigma \vartheta \mu \omega }+13 y^2 \eta ^{\mu \sigma } \varepsilon ^{\vartheta \nu \rho \omega }-19 y z
\eta ^{\nu \rho } \varepsilon ^{\sigma \vartheta \mu \omega }-19 y z \eta ^{\mu \sigma } \varepsilon
^{\vartheta \nu \rho \omega }-4 z^2 \eta ^{\nu \rho } \varepsilon ^{\sigma \vartheta \mu \omega }-4 z^2 \eta
^{\mu \sigma } \varepsilon ^{\vartheta \nu \rho \omega })\,,
\end{eqnarray}
\begin{eqnarray}
&&\langle T \hat{T}_{\bar{\psi}\lambda}^{\mu\nu}(x) \hat{T}_{\bar{\psi}\lambda}^{\sigma\rho}(y) \hat{j}_{A\bar{\psi}\psi}^{\omega}(z) \rangle_c = \frac{4 e^2  e_{\vartheta } (e^{\nu } (e^{\rho } \varepsilon ^{\sigma \vartheta \mu \omega
	}+e^{\sigma } \varepsilon ^{\vartheta \mu \rho \omega })+e^{\mu } (e^{\rho } \varepsilon ^{\sigma
		\vartheta \nu \omega }+e^{\sigma } \varepsilon ^{\vartheta \nu \rho \omega }))}{\pi ^6 (x-y)^3
	(x-z)^4 (y-z)^4}\,,
\end{eqnarray}
\begin{eqnarray}
&&\langle T\, \hat{T}_{\bar{\psi}\lambda}^{\mu\nu}(x) \hat{T}_{\bar{\lambda}\psi}^{\sigma\rho}(y) \hat{j}_{A\bar{\psi}\psi}^{\omega}(z) \rangle_c = \frac{5 }{2 \pi ^6 (x-y)^3 (x-z)^4 (y-z)^4}e_{\vartheta } (-2 e^2 e^{\mu } e^{\rho } \varepsilon ^{\sigma \vartheta \nu \omega }-2
e^2 e^{\nu } (e^{\rho } \varepsilon ^{\sigma \vartheta \mu \omega }+e^{\sigma } \varepsilon
^{\vartheta \mu \rho \omega })\nonumber \\
&&-2 e^2 e^{\mu } e^{\sigma } \varepsilon ^{\vartheta \nu \rho
	\omega }+\eta ^{\nu \rho } \varepsilon ^{\sigma \vartheta \mu \omega }+\eta ^{\mu \rho } \varepsilon ^{\sigma
	\vartheta \nu \omega }+\eta ^{\sigma \nu } \varepsilon ^{\vartheta \mu \rho \omega }+\eta ^{\sigma \mu }
\varepsilon ^{\vartheta \nu \rho \omega })\,.
\label{4}
\end{eqnarray}
\end{widetext}
We see that each of the expressions (\ref{1})-(\ref{4}) separately, does not have the universal conformally symmetric form (\ref{ttj_erdm}). However, when we sum up them in (\ref{9terms}), we get
\begin{eqnarray}
&&\langle T\, \hat{T}^{\mu\nu}(x) \hat{T}^{\sigma\rho}(y) \hat{j}^{\omega}_{A}(z) \rangle_c = -19 \Big(4 \pi^6 (x-y)^5 \nonumber \\
&&\times(x-z)^3
(y-z)^3\Big)^{-1}e_{\vartheta } \Big( \eta ^{\nu \rho } \varepsilon^{\sigma \vartheta \mu \omega }+\eta^{\mu \rho }
\varepsilon^{\sigma \vartheta \nu \omega }\nonumber \\
&& +\eta^{\nu \sigma } \varepsilon^{\vartheta \mu \rho \omega }+\eta^{\mu\sigma } \varepsilon^{\vartheta \nu \rho \omega }  -6 e^2 (e^{\nu } e^{\rho } \varepsilon^{\sigma \vartheta \mu\omega }\nonumber \\
&&+e^{\mu } e^{\rho } \varepsilon^{\sigma \vartheta \nu \omega }+e^{\sigma } e^{\nu } \varepsilon^{\vartheta\mu \rho \omega }  +e^{\sigma } e^{\mu } \varepsilon^{\vartheta \nu \rho \omega })\Big)
\,,
\label{func_main}
\end{eqnarray}
which has the necessary form, following from the symmetries (\ref{ttj_erdm}). The anomaly coefficient will be
\begin{eqnarray}
\mathscr{A}_{RSA} =-19 \mathscr{A}_{s=1/2}=- \frac{19}{\pi^6}
\,.
\label{res_main}
\end{eqnarray}

Taking into account (\ref{Aa}) and (\ref{anom_erdm}), we obtain the following expression for the gravitational chiral anomaly in the RSA theory
\begin{eqnarray}
\langle\nabla_{\mu} \hat{j}^{\mu}_A\rangle_{RSA}=
\frac{-19}{384 \pi^2\sqrt{-g}}\varepsilon^{\mu\nu\rho\sigma}
R_{\mu\nu\kappa\lambda}{R_{\rho\sigma}}^{\kappa\lambda}\,,
\label{anom_main}
\end{eqnarray}
which is --19 times larger compared to the spin 1/2 anomaly (\ref{anom12}).

In this way, we calculated the anomaly and simultaneously verified that the RSA theory, in the infinitely strong coupling limit $ m\to\infty $, satisfies conformal symmetry. Namely, one-loop graphs of the form Fig. \ref{fig} calculated by us in (\ref{func_main}), as well as earlier in (\ref{3point_gauge}) and (\ref{jja_gen1}), meet the conformal symmetry prediction for the form of the three-point functions (\ref{jjj_erdm_gen}) and (\ref{ttj_erdm}).

%================
\section{Discussion}
\label{sec_disc}
%================

%================
\subsection{Interpretation of the factor -19}
\label{subsec_disc_interpret}
%================

Thus, the factor in the gravitational chiral anomaly in the RSA theory (\ref{anom_main}) differs from the well-known value (\ref{anom_old}). The difference between the factors was expected, as it already existed for the gauge chiral anomaly (\ref{anom}), as discussed in \cite{Adler:2017shl}. 

Gauge and gravitational anomalies, e.g. (\ref{anom_old}), (\ref{anom_gen}), for massless higher spin fields were calculated in \cite{Duff:1982yw}. Here the particles of spin S are defined as having on mass shell two polarization states with chirality $ \pm $ S. If we switch off the interaction the Adler model describes on mass shell one field of spin 3/2 and two fields of spin 1/2. The same counting should apply when we keep interaction non-vanishing but tend momentum to the infinity. This limit is relevant to evaluation of the chiral anomaly. In this way the coefficient $ -19 $ is readily restored from (\ref{anom_old}) as
\begin{eqnarray}
-19=-21+2\,. 
\end{eqnarray} 

A similar conclusion can be reached if we analyze the contribution of the ghosts. In particular, the factor $ -21 $ in (\ref{anom_old}) can be obtained as $-21=-20-1$, where $-20$ is the ``ghostless'' contribution and $ -1 $ is the contribution of the ghosts \cite{Carrasco:2013ypa}.

Now, the $-19$ factor could be obtained by adding the ghostless contribution and the contribution of spin 1/2, i.e.
\begin{eqnarray}
-19=-20+1\,.
\end{eqnarray}

Note that, as discussed in \cite{Adler:2017shl}, a similar correspondence is also observed for the gauge chiral anomaly, for which $5=4+1$, with 4 being the ghostless contribution, and 1 being the contribution from spin 1/2 field. The need to consider the ghostless part of the anomaly can be motivated by the fact that ghosts do not propagate in the RSA theory, and the need to add the contribution of the field with spin 1/2 corresponds to the additional field $ \lambda $.
%
%Of course, this reasoning above cannot be considered as a rigorous derivation of (\ref{anom_main}). It is enough to notice, that the $ \lambda $ field cannot be considered free, since it interacts infinitely strongly with the $ \psi_{\mu} $ field, and for any $ m $ the propagator $ \langle\lambda\bar{\lambda} \rangle$ is equal to zero and, of course, we do not quantize gravitational field. At least, this indicates the absence of a fundamental discrepancy between the obtained anomaly factor and the factors in other theories.

%================
\subsection{Numerical factors and Landau levels}
%================

Another aspect of appearing numerical factors may be discussed in connection with 
the Landau levels for spin $3/2$ particles \cite{dePaoli:2012eq}
which was recently explored \cite{Dexheimer:2021sxs} in the studies 
of Delta baryons in strongly magnetized neutron-star matter.

The results of \cite{dePaoli:2012eq} may be interpreted as pointing out to  the gyromagnetic ratio $g=2$ for 
spin $3/2$ particles  corresponding in fact to its ``natural'' value \cite{Ferrara:1992yc}. 

Passing to relativistic case by substitution  $E \to (E^2-m^2)/2m$ 
(see e.g. \cite{Dokshitzer:2021aji}), one may cancel mass $m$ in the denominator with the one in Bohr magneton and 
approach the chiral limit $m=0$. In that case the zero mode is obtained 
by cancellation of spin energy with the orbital one at the first excited Landau level \footnote{The tachyon mode from LLL requires the separate investigation and may be hopefully eliminated in the framework of Adler procedure.}. Treating the Landau levels flow similarly to the one for spin $1/2$, the ratio $5$ in (\ref{anom}) can be obtained  
by adding to the $3/2$ contribution the one from $1/2$ multiplied by two\footnote{The $1/2$ contribution is provided by Adler procedure with the proper treatment of ghosts, as discussed in previous subsection,  which may have also counterparts in \cite{dePaoli:2012eq}.}
\begin{eqnarray}
5= \frac{\frac{3}{2}+ 2 \cdot \frac{1}{2}}{ \frac{1}{2}}
\label{LL5}\,.
\end{eqnarray}

Factor $-19$ may in fact be represented in a similar way 
\begin{eqnarray}
-19= \frac{\phi(\frac{3}{2})+ 2 \cdot \phi(\frac{1}{2})}{\phi(\frac{1}{2})}\,,
\label{LL19}\,
\end{eqnarray}
with $\phi(S) = S-2 S^3$ from (\ref{anom_gen}). The interpretation in terms of Landau levels flow is not directly applicable here, as the spin precession frequency in gravitational field is fixed by equivalence principle to be equal to that 
of orbital angular momentum (see \cite{Teryaev:2016edw} and Ref. therein), while the r.h.s. of  (\ref{anom_gen})
is not controlled by equivalence principle containing the $S^3$ term.

%------------------------------------------------------------------------------------

%================
\section{Conclusion}
\label{sec_concl}
%================

We have calculated the gravitational chiral anomaly for the extended Rarita-Schwinger-Adler model for spin $3/2$ fields and found that it is $-19$ times larger than the standard anomaly for spin 1/2. The obtained factor differs from the well-known factor $-21$, for the Rarita-Schwinger field theory. We can associate the $ -19 $ factor with the results for spin 3/2 from supergravity \cite{Duff:1982yw} for free fields, by counting the degrees of freedom and their contributions. Then it is obtained either as $ -19=-21+2 $ or $ -19=-20+1 $.

We have also calculated the chiral gauge anomaly in RSA theory. The obtained numerical factor coincides with the original result obtained using the shift method \cite{Adler:2017shl}.

At the same time, our derivation of the anomalies is a direct verification of the conformality of the RSA theory in the limit of the infinitely strong interaction $ m\to\infty $ at the level of one-loop three-point functions. We have explicitly shown, that one-loop three-point functions with two vector currents and one axial current, as well as with two stress-energy tensors and one axial current, satisfy the consequences of conformal symmetry \cite{Erdmenger:1996yc, Erdmenger:1999xx}.

The method used by us for calculating the gravitational chiral anomaly does not need a transition to a curved space-time. The calculation of a three-point function consists in a simple multiplication of usual propagators in the coordinate space.

The obtained result can be used when considering the theories beyond the Standard Model, where the fields with higher spins should participate in the cancellation of anomalies. Another application relates to the cutting edge field of manifestations of quantum anomalies in the properties of relativistic fluids and in condensed matter, where a number of new effects corresponding to various quantum anomalies have been discovered. However, consideration of these applications is beyond the scope of this work.

\appendix

%================
\section{Three-point function with arbitrary points}
\label{append}
%================

In this section, we will verify that the three-point function with currents $ \langle \hat{j}_V(x)\hat{j}_V(y)\hat{j}_A(z) \rangle $ satisfies the general conformally symmetric form for an arbitrary position of the points $ x_{\mu}, y_{\nu}, z_{\alpha} $ and once again check the factor from the anomaly (\ref{anom}).

Without loss of generality, we can put $ z =0 $, since both sides in (\ref{jjj_erdm_gen}), obviously, have translation invariance with respect to the simultaneous shift of all three variables. As before, the three-point function is described by a single term (\ref{corr_gauge}), which now, however, must be calculated using the general form of the propagator (\ref{prop_x}) (in which, however, we can again neglect the terms of the order $ 1/m^2 $).

The correlator is calculated according to the same algorithm as in the Section  \ref{subsec_calc_gauge}. Finally, we obtain
\begin{eqnarray} 
&&\langle T \hat{j}_V^{\mu}(x) \hat{j}_V^{\nu}(y) \hat{j}^{\omega}_{A}(0) \rangle_c =  \frac{5 }{\pi^6 x^4 y^4 (x-y)^4}\Big(
y^2 x_{\vartheta}\varepsilon^{\mu\nu w \vartheta}
\nonumber \\
&&-x^2 y_{\vartheta}\varepsilon^{\mu\nu w \vartheta}
-2 y^{\nu} x_{\vartheta}y_{\eta} \varepsilon^{\mu w \vartheta \eta}+2 x^{\mu} x_{\vartheta}y_{\eta} \varepsilon^{\mu w \vartheta \eta}\Big)\,.\qquad
\label{jja_gen1}
\end{eqnarray}
On the other hand, from (\ref{jjj_erdm_gen}) when $ z=0 $ we obtain
\begin{eqnarray} 
&&\langle T \hat{j}_V^{\mu}(x) \hat{j}_V^{\nu}(y) \hat{j}^{\omega}_{A}(z) \rangle_c =  \frac{4 \mathscr{B}}{x^4 y^4 (x-y)^4}\Big(
y^2 x_{\vartheta}\varepsilon^{\mu\nu w \vartheta}
\nonumber \\
&&-x^2 y_{\vartheta}\varepsilon^{\mu\nu w \vartheta}
-2 y^{\nu} x_{\vartheta}y_{\eta} \varepsilon^{\mu w \vartheta \eta}+2 x^{\mu} x_{\vartheta}y_{\eta} \varepsilon^{\mu w \vartheta \eta}\Big)\,.\qquad
\label{jja_gen2}
\end{eqnarray}
Comparing (\ref{jja_gen1}) and (\ref{jja_gen2}), we see that the three-point function has a conformally symmetric form and the factor in the anomaly is the same as in the Section \ref{subsec_calc_gauge}
\begin{eqnarray}
\mathscr{B}=\frac{5}{4\pi^6}\,,
\label{b_rsa2}
\end{eqnarray}
and thus the anomaly (\ref{anom}) is confirmed.

%=================================================
{\bf Acknowledgments}
%=================================================

GP is thankful to A. F. Pikelner and A. A. Golubtsova for valuable discussions.
The work was supported by Russian Science Foundation Grant No. 21-12-00237, the work of VIZ is partially supported by grant No. 0657-2020-0015 of the Ministry of Science and Higher Education of Russia.

\bibliography{lit}

\begin{thebibliography}{10}

\bibitem{Adler:2017shl}
Stephen~L. Adler.
\newblock {Analysis of a gauged model with a spin-$\frac{1}{2}$ field directly
  coupled to a Rarita-Schwinger spin-$\frac{3}{2}$ field}.
\newblock {\em Phys. Rev. D}, 97(4):045014, 2018.

\bibitem{Adler:2019zxx}
Stephen~L. Adler and Pablo Pais.
\newblock {Chiral anomaly calculation in the extended coupled Rarita-Schwinger
  model}.
\newblock {\em Phys. Rev. D}, 99(9):095037, 2019.

\bibitem{Adler:2015yha}
Stephen~L. Adler.
\newblock {Classical Gauged Massless Rarita-Schwinger Fields}.
\newblock {\em Phys. Rev. D}, 92(8):085022, 2015.

\bibitem{Duff:1982yw}
M.~J. Duff.
\newblock {Ultraviolet divergences in extended supergravity}.
\newblock In {\em {First School on Supergravity}}, 1 1982.

\bibitem{Adler:2019exo}
Stephen~L. Adler.
\newblock {Analysis of an $SU(8)$ model with a spin-$\frac{1}{2}$ field
  directly coupled to a gauged Rarita\textendash{}Schwinger spin-$\frac{3}{2}$
  field}.
\newblock {\em Int. J. Mod. Phys. A}, 34(33):1950230, 2019.

\bibitem{Carrasco:2013ypa}
J.~J.~M. Carrasco, R.~Kallosh, R.~Roiban, and A.~A. Tseytlin.
\newblock {On the U(1) duality anomaly and the S-matrix of N=4 supergravity}.
\newblock {\em JHEP}, 07:029, 2013.

\bibitem{Nielsen:1978ex}
N.~K. Nielsen, Marcus~T. Grisaru, H.~Romer, and P.~van Nieuwenhuizen.
\newblock {Approaches to the Gravitational Spin 3/2 Axial Anomaly}.
\newblock {\em Nucl. Phys. B}, 140:477--498, 1978.

\bibitem{Erdmenger:1996yc}
J.~Erdmenger and H.~Osborn.
\newblock {Conserved currents and the energy momentum tensor in conformally
  invariant theories for general dimensions}.
\newblock {\em Nucl. Phys. B}, 483:431--474, 1997.

\bibitem{Erdmenger:1999xx}
Johanna Erdmenger.
\newblock {Gravitational axial anomaly for four-dimensional conformal field
  theories}.
\newblock {\em Nucl. Phys. B}, 562:315--329, 1999.

\bibitem{Chernodub:2021nff}
Maxim~N. Chernodub, Yago Ferreiros, Adolfo~G. Grushin, Karl Landsteiner, and
  Mar\'\i{}a A.~H. Vozmediano.
\newblock {Thermal transport, geometry, and anomalies}.
\newblock 10 2021.

\bibitem{Son:2009tf}
Dam~T. Son and Piotr Surowka.
\newblock {Hydrodynamics with Triangle Anomalies}.
\newblock {\em Phys. Rev. Lett.}, 103:191601, 2009.

\bibitem{Zakharov:2012vv}
Valentin~I. Zakharov.
\newblock {Chiral Magnetic Effect in Hydrodynamic Approximation}.
\newblock 2012.
\newblock [Lect. Notes Phys.871,295(2013)].

\bibitem{Kharzeev:2013jha}
Dmitri Kharzeev, Karl Landsteiner, Andreas Schmitt, and Ho-Ung Yee.
\newblock {Strongly Interacting Matter in Magnetic Fields}.
\newblock {\em Lect. Notes Phys.}, 871:pp.1--624, 2013.

\bibitem{Sadofyev:2010is}
A.~V. Sadofyev, V.~I. Shevchenko, and V.~I. Zakharov.
\newblock {Notes on chiral hydrodynamics within effective theory approach}.
\newblock {\em Phys. Rev.}, D83:105025, 2011.

\bibitem{Prokhorov:2021bbv}
G.~Yu. Prokhorov, O.~V. Teryaev, and V.~I. Zakharov.
\newblock {Chiral Vortical Effect in Extended Rarita-Schwinger Field Theory and
  Chiral Anomaly}.
\newblock 9 2021.

\bibitem{Khaidukov:2020mrn}
Z.~V. Khaidukov and R.~A. Abramchuk.
\newblock {Chiral separation effect for spin 3/2 fermions}.
\newblock {\em JHEP}, 07:183, 2021.

\bibitem{Stone:2018zel}
Michael Stone and Jiyoung Kim.
\newblock {Mixed Anomalies: Chiral Vortical Effect and the Sommerfeld
  Expansion}.
\newblock {\em Phys. Rev.}, D98(2):025012, 2018.

\bibitem{Landsteiner:2011cp}
Karl Landsteiner, Eugenio Megias, and Francisco Pena-Benitez.
\newblock {Gravitational Anomaly and Transport}.
\newblock {\em Phys. Rev. Lett.}, 107:021601, 2011.

\bibitem{Prokhorov:2020okl}
G.~Yu. Prokhorov, O.~V. Teryaev, and V.~I. Zakharov.
\newblock {Chiral vortical effect: Black-hole versus flat-space derivation}.
\newblock {\em Phys. Rev. D}, 102(12):121702(R), 2020.

\bibitem{Prokhorov:2020npf}
G.~Yu Prokhorov, O.~V. Teryaev, and V.~I. Zakharov.
\newblock {Chiral vortical effect for vector fields}.
\newblock {\em Phys. Rev. D}, 103(8):085003, 2021.

\bibitem{Robinson:2005pd}
Sean~P. Robinson and Frank Wilczek.
\newblock {A Relationship between Hawking radiation and gravitational
  anomalies}.
\newblock {\em Phys. Rev. Lett.}, 95:011303, 2005.

\bibitem{Patel:2015tea}
Hiren~H. Patel.
\newblock {Package-X: A Mathematica package for the analytic calculation of
  one-loop integrals}.
\newblock {\em Comput. Phys. Commun.}, 197:276--290, 2015.

\bibitem{Shapiro:2016pfm}
Ilya~L. Shapiro.
\newblock {Covariant derivative of fermions and all that}.
\newblock 11 2016.

\bibitem{Das:1978tm}
Ashok~K. Das.
\newblock {On the stress tensor in a class of gauge theories}.
\newblock {\em Phys. Rev. D}, 18:2065, 1978.

\bibitem{Grozin:2003ak}
Andrey~G. Grozin.
\newblock {Lectures on multiloop calculations}.
\newblock {\em Int. J. Mod. Phys. A}, 19:473--520, 2004.

\bibitem{Kazakov:2008tr}
D.~I. Kazakov.
\newblock {Radiative Corrections, Divergences, Regularization, Renormalization,
  Renormalization Group and All That in Examples in Quantum Field Theory}.
\newblock 2008.

\bibitem{Nakayama:2013is}
Yu~Nakayama.
\newblock {Scale invariance vs conformal invariance}.
\newblock {\em Phys. Rept.}, 569:1--93, 2015.

\bibitem{dePaoli:2012eq}
M.~G. de~Paoli, D.~P. Menezes, L.~B. Castro, and C.~C. Barros, Jr.
\newblock {The Rarita-Schwinger Particles Under de Influence of Strong Magnetic
  Fields}.
\newblock {\em J. Phys. G}, 40:055007, 2013.

\bibitem{Dexheimer:2021sxs}
Veronica Dexheimer, Kauan~D. Marquez, and D\'ebora~P. Menezes.
\newblock {Delta Baryons in Neutron-Star Matter under Strong Magnetic Fields}.
\newblock {\em Eur. Phys. J. A}, 57:216, 2021.

\bibitem{Ferrara:1992yc}
Sergio Ferrara, Massimo Porrati, and Valentine~L. Telegdi.
\newblock {$g=2$ as the natural value of the tree-level gyromagnetic ratio of
  elementary particles}.
\newblock {\em Phys. Rev. D}, 46:3529--3537, 1992.

\bibitem{Dokshitzer:2021aji}
O.~V. Teryaev.
\newblock {\em {Axial anomaly as landau levels flow, decoupling and monopole
  pairs production, Gribov-90 Memorial Volume}}, pages 509--515.

\bibitem{Teryaev:2016edw}
O.~V. Teryaev.
\newblock {Gravitational form factors and nucleon spin structure}.
\newblock {\em Front. Phys.(Beijing)}, 11(5):111207, 2016.

\end{thebibliography}

\end{document}